\documentclass[%
 reprint,
 superscriptaddress,
 amsmath,amssymb,
prb,
]{revtex4-2}

\usepackage{graphicx}
\usepackage{dcolumn}
\usepackage{bm}
\usepackage{textcomp}
\usepackage{subcaption}
\usepackage{color}
\usepackage{soul}

\begin{document}

\preprint{APS/123-QED}

\title{Evolution of ferromagnetic stripes in FePt thin films at low temperature}%

\author{Cynthia P. Quinteros}
 \email{cquinteros@unsam.edu.ar}
\affiliation{%
 ECyT-UNSAM, CONICET, Martín de Irigoyen 3100, B1650JKA, San Martín, Bs. As., Argentina.
}%

\author{Dafne Goijman}
\affiliation{%
Instituto de Nanociencia y Nanotecnología, INN (CNEA-CONICET). Centro Atómico Bariloche, R4802AGP, Río Negro, Argentina. 
}
\affiliation{%
 UNRN, Sede Andina, R8400GNA, Río Negro, Argentina. 
}%

\author{Silvia Damerio}
 \altaffiliation[Currently at ]{Institut de Ciència de Materials de Barcelona (ICMAB-CSIC), Campus de la UAB, Bellaterra, 08193, Spain.}
\affiliation{
 Zernike Institute for Advanced Materials, University of
Groningen, 9747 AG Groningen, The Netherlands.
}%


\author{Julián Milano}
\affiliation{%
Instituto de Nanociencia y Nanotecnología, INN (CNEA-CONICET). Centro Atómico Bariloche, R4802AGP, Río Negro, Argentina. 
}
\affiliation{%
 Instituto Balseiro, UNCuyo-CNEA, R4802AGP, Río Negro,
Argentina.
}





\date{\today}

\begin{abstract}
\noindent Patterns of ferroic domains and domain walls are being intensively studied to implement new logic schemes. Any technological application of such objects depends on a detailed understanding of them. This study analyzes patterns of ferromagnetic stripes on equiatomic FePt thin films at low temperatures. Since FePt is known to develop a transition from in-plane homogeneous magnetization to stripes upon varying its thickness, multiple samples are studied to consider the critical value within the analyzed range. Stripes’ width demonstrates the well-known Murayama‘s law while a non-trivial dependence on temperature is also reported. Moreover, the room-temperature uniform distribution of the pattern evolves into a distorted one upon temperature cycling. Finally, dissimilar striped patterns are obtained upon reducing and increasing temperature indicating the states are dependent on the history of applied stimuli rather than the parametric conditions.   
\end{abstract}

\maketitle


\section*{\label{sec:Intro}Introduction}


\noindent Textures of the order parameter spontaneously develop within ferroic materials to minimize their internal energy \cite{kittel_theory_1946,williams_magnetic_1957,saito_new_1964,murayama_micromagnetics_1966}. 
Recently, those usually complex and intricated spatial distributions are being studied in terms of their resemblance of the \textit{in-materio} implementation of networks for 'neuromorphic-like' novel computing strategies \cite{rieck_ferroelastic_nodate}. 
While the zones where the order parameter is homogeneous (domains) display electrical properties ascribed to the bulk behavior (i.e. dielectric or insulating nature), their boundaries (domain walls) have shown a plethora of diverse and promising characteristics, among which conduction is perhaps the most attractive one \cite{salje_multiferroic_2010,catalan_domain_2012}. The idea of exploiting a self-formed matrix of insulating islands connected and/or surrounded by multiple conductive paths, both suitable for resistive switching, is an appealing strategy to implement physical realizations of highly interconnected switching hubs. Compared to crossbar arrays, in which the packing density relies so dramatically on the lithographic tool \cite{likharev_hybrid_2007}, the possibility of driving self-assemblies to implement useful computational tasks is a very attractive concept that is being intensively explored nowadays \cite{feigl_controlled_2014,nahas_inverse_2020}.

Fe$_{0.5}$Pt$_{0.5}$ (FePt) is a ferromagnetic (FM) metal alloy with a rich phase diagram \cite{cahn_binary_1991}. Sputtered at room temperature, the so-called A1 phase, FCC crystalline-disordered soft-magnetic structure, is obtained \cite{Vasquez2008}. Above a certain critical thickness ($t_{crit}$), FePt films form a striped magnetic pattern \cite{sallica_leva_magnetic_2010}. This corresponds to elongated domains with both in-plane and out-of-plane components of the magnetization in which the in-plane component lies along the stripes 
\cite{martinez_modeling_2016}. Due to finite size considerations, closure domains consisting of in-plane components perpendicular to the stripes may also form to reduce the stray magnetic field \cite{fin_-plane_2015}. 

The FM domain pattern and the type of magnetic interactions acting on the film are strongly correlated, that is, the planar domains correspond to positive interactions (exchange) while negative interactions, such as dipolar, modulate the energy landscape \cite{alvarez_correlation_2014}. Since the theory of stripe domains was developed \cite{kittel_theory_1946}, theoretical and numerical approaches have been applied to understand and predict the magnetic properties of the pattern \cite{saito_new_1964,murayama_micromagnetics_1966}. According to those methods, based on minimizing the magnetic free energy, all the properties of the stripe domains can be modeled by taking into account the film magnetization and thickness, the perpendicular magnetic anisotropy, $K_{\rm PMA}$, and the exchange stiffness, $A$. However, regardless of the growth conditions, experimental thin films differ from modeled entities due to the appearance of unavoidable topographical defects and other sources of inhomogeneity. 
These unconsidered aspects give rise to metastable states which are not fully understood. 

%
For practical applications that rely on self-assembled networks arising from magnetic patterns, a comprehensive model of the domain evolution is required. A detailed understanding of the width and variability of the stripes as well as their rectitude and connectivity, depending on external conditions could, in turn, be used to engineer the texture on demand. 

In this work, we demonstrate the evolution of the magnetic stripes with temperature (T), by studying an archetypical ferromagnetic material (FePt). %
The strategy of following the magnetic domains in films of different thicknesses at low T enables the direct visualization of the pattern evolution. Moreover, we present evidence indicating that the domain configuration depends not only on the parametric conditions (i.e. the absolute T) but on the thermal history. 

\section*{\label{sec:Exp}Experimental section}

\begin{figure}
\includegraphics[width=0.9\columnwidth]{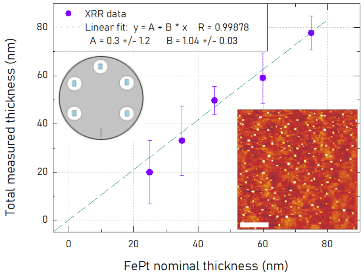}
\caption{\label{fig:XRR}XRR-determined total thickness (FePt films covered by a Ru capping layer) as a function of the nominal FePt thickness. The experimental data are fitted using a linear function close to the identity. A sketch of the substrates' positions on the carrousel holder is included as an inset. On the bottom right, an AFM map of the bare substrate is displayed. The scale bar is 2 $\mu$m-long.} 
\end{figure}

\noindent FePt films were deposited by DC magnetron sputtering on regular silicon wafers. Pieces of 5$\times$5~mm$^2$ size, cut from the same semiconducting Si wafer, were used as substrates. A cleaning procedure was conducted beforehand although the persistent formation of a thin native oxide layer (SiO$_2$) is expected. Bottom-right inset of Fig. \ref{fig:XRR} comprises an atomic force microscopy (AFM) of the bare substrate. Films of different thicknesses ranging from 25~nm to 75~nm were grown. A 3~nm-thick Ru capping layer was deposited to preserve the FM films from oxidation. Although nominally equiatomic, the commercial FePt target used in this deposition has a Fe atomic percent composition close to 45\% \cite{Vasquez2008}. FePt cathode was sputtered at 20~W (power density 1.8~W/cm$^2$) in a 2.6~mTorr Ar atmosphere resulting in a deposition rate of $\sim$~0.16~nm/s.

By tuning the time each piece of substrate is exposed to the plume, FePt films of different thicknesses were deposited sequentially. All of them were loaded simultaneously in the sputtering chamber, distributed in different angular positions of a carrousel holder (see top-left inset of Fig. \ref{fig:XRR}). This strategy reduces the occurrence of unexpected variations while loading the chamber multiple times (see Appendix of \cite{quinteros_impact_2020}). Fig. \ref{fig:XRR} shows the measured thickness, $t_{tot}^{measured}$, using X-ray Reflectivity (XRR) as a function of FePt nominal values, $t_{FePt}^{nominal}$. The measured values (corresponding to the sum of FePt and Ru layers) are fitted using a linear function close to the identity. This accounts for a small overestimation of the deposition rates indicating that the actual values are slightly lower than expected. Samples' thicknesses and designations are summarized in Table \ref{tab1}.

\begin{table}[htbp]
\caption{Samples designation along with their nominal FePt thicknesses, $t_{FePt}^{nominal}$, and the experimentally determined total values, $t_{tot}^{measured}$ (as depicted in Fig. \ref{fig:XRR}).}
\begin{center}
\begin{tabular}{|c|c|c|}
\hline
\textbf{\textit{Sample}} & $t_{FePt}^{nominal}$ [nm] & $t_{tot}^{measured}$ + $\Delta t_{tot}^{measured}$ [nm] \\
\hline
\textbf{F75} & 75 & 78 $\pm$ 7\\
\hline
\textbf{F60} & 60 & 59 $\pm$ 10\\
\hline
\textbf{F45} & 45 & 50 $\pm$ 6\\
\hline
\textbf{F35} & 35 & 33 $\pm$ 14\\
\hline
\textbf{F25} & 25 & 20 $\pm$ 13\\
\hline
\end{tabular}
\label{tab1}
\end{center}
\end{table}

\noindent The magnetic properties were studied by means of SQUID magnetometry (Quantum Design MPMS-XL 7) in a range of temperatures varying from 10 K to 300 K and at in-plane fields up to 2 T.

The magnetic force microscopy (MFM) experiments presented in this study were performed with a customized Attocube\textsuperscript{\textregistered} scanning probe microscope inserted in a Quantum Design Physical Property Measurement System (PPMS). Multiple scans were collected at different temperatures ranging from 300 K down to 10 K. No field was intentionally applied during the scans. The sample surface was scanned using commercial (Nanoworld) CoCr-coated tips. The images were collected in dual-pass tapping mode, with a second scan lift of 10 nm. The data were then processed with the open-source software Gwyddion \cite{necas_gwyddion_2012}.

\section*{\label{sec:Results}Results}

\begin{figure*}[t!]
    \centering
    \begin{subfigure}[b]{0.5\textwidth}
        \centering
        \includegraphics{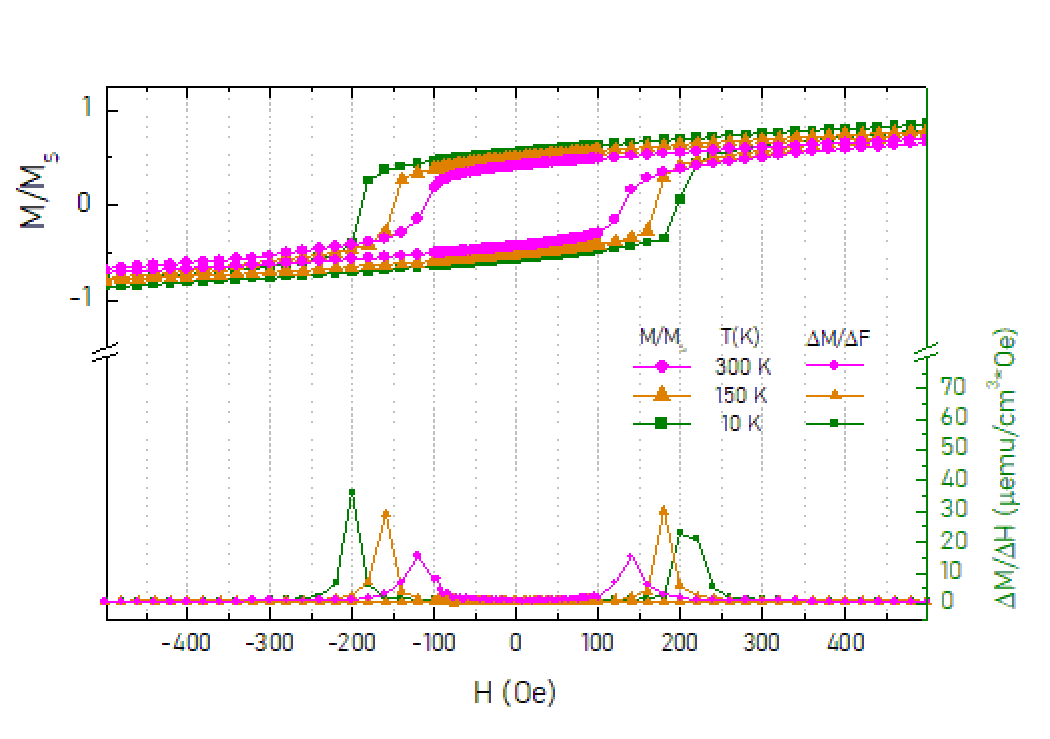}
        \caption{\label{fig:MH_F1s_F3s_a}}
    \end{subfigure}%
    ~ 
    \begin{subfigure}[b]{0.5\textwidth}
        \centering
        \includegraphics{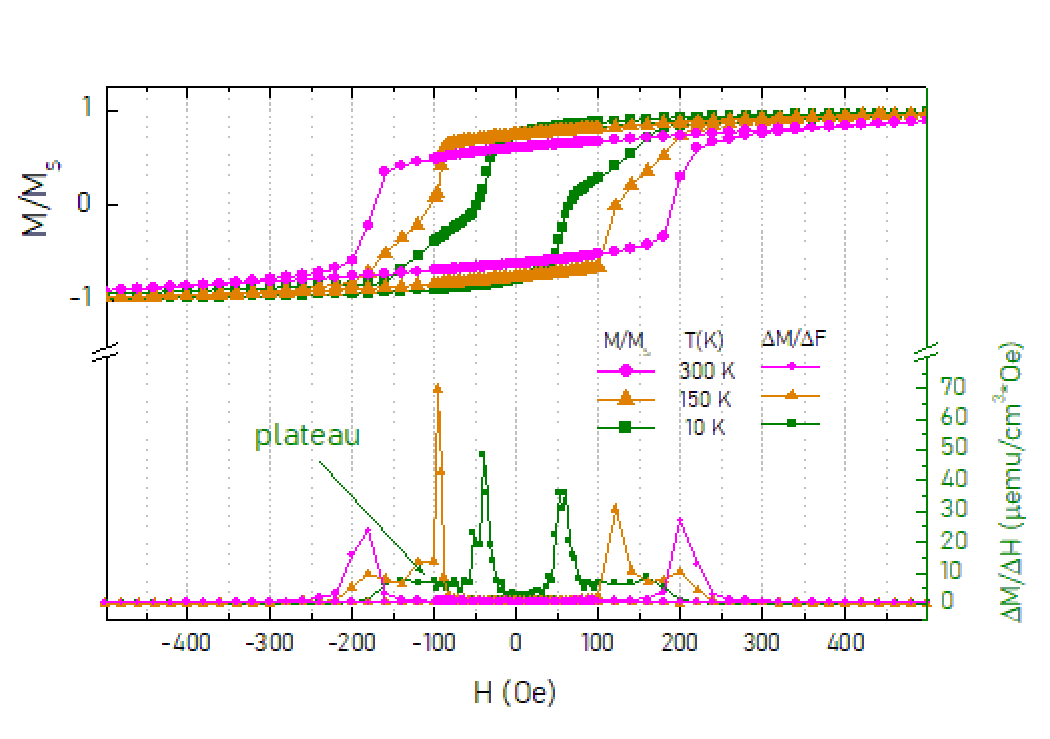}
        \caption{\label{fig:MH_F1s_F3s_b}}
    \end{subfigure}
    \caption{\label{fig:MH_F1s_F3s}Normalized in-plane magnetization ($\frac{M}{M_{sat}}$) as a function of the magnetic field ($H$) measured on (a) \textbf{F75} and (b) \textbf{F45} samples quantified at three different temperatures ($T$ = 10, 150, and 300 K). In both cases, the right axis displays the rate of change of $M$ with the variation of $H$ ($\frac{\Delta M}{\Delta H}$). In (b), measurements recorded at 150 K and 10 K display plateaux accompanying the peaks associated with their corresponding coercive conditions.}
\end{figure*}

\noindent Macroscopic measurements, such as the magnetization as a function of the magnetic field, $M-H$ (see Fig. S1 in the Supplementary Information), demonstrate the same dependencies as reported for A1 FePt films elsewhere  \cite{Vasquez2008,sallica_leva_magnetic_2010,guzman_abnormal_2013,martinez_modeling_2016}. Depending on the film thickness, there are two magnetic regimes: fully in-plane homogeneous magnetization and striped pattern. The threshold between them is a critical thickness, $t_{crit}$, which is determined by the temperature (T) and the specific deposition conditions \cite{guzman_abnormal_2013,alvarez_tunable_2015}. At RT, $t_{crit} \sim$ 30~nm \cite{Vasquez2008}. Thinner films are expected to display reduced coercive ($H_{C}$) and saturation ($H_{sat}$) fields (see Fig. S2 and S3 in the Supplementary Information) \cite{sallica_leva_magnetic_2010}. In contrast, films thicker than $t_{crit}$ demonstrate increased $H_{C}$ and $H_{sat}$ values which, in turn, depend on the film details (see Fig. S2 and S3) \cite{sallica_leva_magnetic_2010}. 
These films display a striped pattern due to the coexistence of in-plane and out-of-plane favorable free-magnetization energy terms. Their relative intensity and, consequently, the balance between them 
depend on the film thickness \cite{sallica_leva_magnetic_2010}, 
the residual tension 
\cite{alvarez_tunable_2015,alvarez_critical_2016}, 
and T \cite{guzman_abnormal_2013}, among other factors. 

\begin{figure*}
\includegraphics[width=\textwidth]{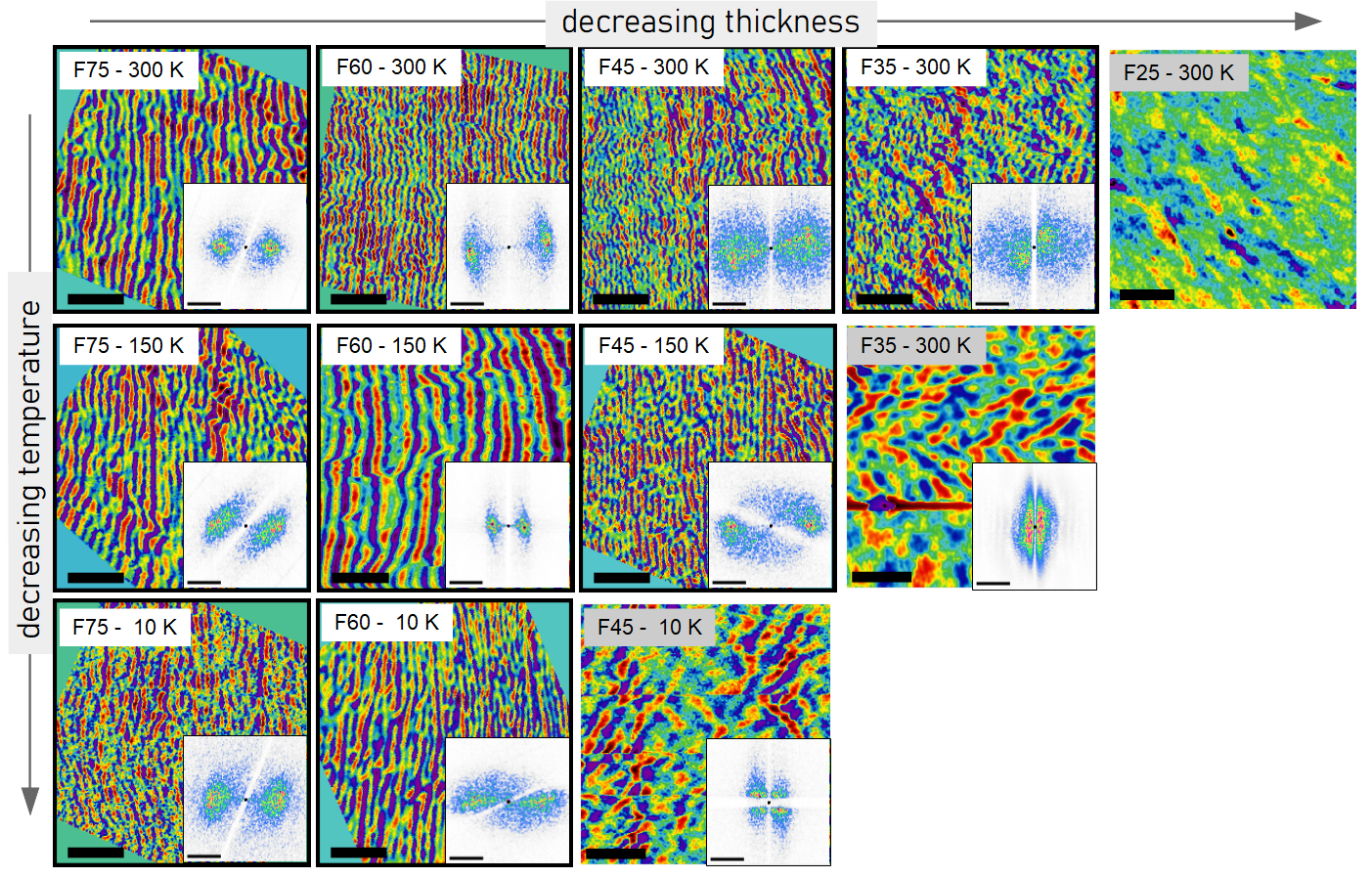}
\caption{\label{fig:MFM-images}Magnetic Force Microscopy images recorded on samples of different thicknesses at different temperatures (T = 10, 150, and 300 K). The bold frames indicate the conditions in which the stripes are clearly visible. The scale bar is 1 $\mu$m long. Insets: 2D FFT for each corresponding MFM. Scale bar is 5 $\mu$m$^{-1}$ long.} 
\end{figure*}

The left axis in Figure \ref{fig:MH_F1s_F3s} shows the normalized in-plane magnetization ($\frac{M}{M_{sat}}$) as a function of the externally applied magnetic field ($H$) measured on two FePt films: \textbf{F75} and \textbf{F45}, respectively. $M$-$H$ loops were recorded for every film at three different temperatures, T = 300, 150, and 10 K (see Fig. S1). In Fig. \ref{fig:MH_F1s_F3s_a}, \textbf{F75} displays an increased coercive field ($H_C$) and remnant magnetization ($M_{rem}$) upon T reduction. This is explained as an increase of the in-plane magnetization to the detriment of the out-of-plane component. Since the magnetization reversal is thermally assisted, an increase of $H_C$ is easily understandable while lowering T. In Fig. \ref{fig:MH_F1s_F3s_b}, upon T reduction, \textbf{F45} also demonstrates an increase in $M_{rem}$ but a concomitant reduction of $H_C$. Although it may seem counterintuitive, this abnormal $H_C-T$ dependence has been reported before \cite{guzman_abnormal_2013,sharma_anomalous_2011}. Associated with the microstructure \cite{sharma_anomalous_2011}, this dependency was pointed as an indication of the change in the magnetic texture \cite{guzman_abnormal_2013}. 
%
%
By differentiating $\frac{M}{M_{sat}}$ with respect to $H$ ($\frac{\Delta M}{\Delta H}$), a measure of the associated rate of change is obtained \cite{berger_magnetization_1996}. Such analysis allows us to properly define $H_C$ as the field at which $\frac{\Delta M}{\Delta H}$ becomes maximum. Moreover, it comprises a tool to gain more insight into how the coercive condition is reached. Noteworthy, the measurements performed on \textbf{F45} at T = 150 and 10 K demonstrate the presence of plateaux. In this case, upon lowering T, not only a reduction of $H_C$ is observed but the shape around the coercive condition gets distorted. This can also be seen in the $\frac{\Delta M}{\Delta H}$-$H$ dependency. While at RT, $\frac{\Delta M}{\Delta H}$ describes a peak around the coercive field, at low T, $\frac{\Delta M}{\Delta H}$ a maximum followed by a plateau is observed for each polarity. A fixed value for the derivative $\frac{\Delta M}{\Delta H}$ accounts for a sustained rate of change of \textit{M} until $H_{sat}$ is reached. The observation of two clearly distinguishable behaviors in the $M-H$ and $\frac{\Delta M}{\Delta H}$-$H$ dependencies of \textbf{F45} reveals the coexistence of two mechanisms for the magnetization reversal. This seems to be related to the change in the stripes as observed in Fig. \ref{fig:MFM-images}. 

The balance between the in-plane and out-of-plane favorable terms gets particularly affected by T, implying that $t_{crit} \equiv f(\mathrm{T})$. In particular, lowering T increases $t_{crit}$. As a consequence, films whose thicknesses at RT are beyond $t_{crit}(\mathrm{RT})$ display a striped pattern that could become fully homogeneously in-plane magnetized at lower T. To directly visualize this transition, Fig. \ref{fig:MFM-images} consists of the MFM images (phase signal) acquired for every sample at three different T (300, 150, and 10 K). The bottom-right insets represent the 2D Fast Fourier Transform (FFT) of each image. Arranged as a set of lines and columns, each line of Fig. \ref{fig:MFM-images} accounts for the dependence of the stripes' width as a function of FePt thickness at a fixed T (see Fig. S4). Consistent with $t_{crit}(RT)$ $\sim$ 30 nm \cite{Vasquez2008}, at RT, four of the five samples display stripes ($t_{\textbf{F25}}$ $<$ 30 nm). As predicted by Murayama \cite{murayama_micromagnetics_1966}, their periodicities get monotonously reduced when the films become thinner. Upon reduction of T, $t_{crit}$ increases, as can be deduced from the fact that one additional film falls below the critical value and its stripes are not clearly visible anymore. At 150 K, \textbf{F35} does not longer display stripes while at 10 K, \textbf{F45} cannot be considered to show such a pattern. Noteworthy, the stripes of those films well above the critical thicknesses for the whole T-range explored here (\textbf{F75} and \textbf{F60}), also change upon varying T. This is consistent with Fig. \ref{fig:MH_F1s_F3s_a} and arises from the variation suffered by the free energy terms whose competition determines the stripes formation. 
Except for \textbf{F60} at 150 K, which will be discussed in the following, the width of the stripes seems to widen upon T reduction (see the trend in Fig. S4 and the 2D FFT's profiles in Fig. S5), indicating a reduction in the number of domain walls. This is consistent with the decrease of the stripe period with increasing T, for T $\rightarrow$ $T_C$ (Curie temperature) $>>$ RT, reported in theoretical studies \cite{barker_dipolar_1983,seul_evolution_1992}. The reason for this counterintuitive behavior relies on the predominance of the dipolar term due to the scaling with T of the competing energy terms \cite{seul_evolution_1992}. 

Nevertheless, the pattern displayed by \textbf{F60} at 150 K does not seem to follow either the trend expected as a function of the thickness (along the line, see Fig. S4) or the one obtained at different T (along the column). This specific image was acquired at 150 K after cooling down to 10 K. So far, we were not indicating this difference considering the parametric condition (i.e. the absolute T at which the MFM image was acquired) should be enough to characterize the state. However, this observation leads us to further explore the importance of the thermal history of the samples.              

\begin{figure}
\includegraphics[width=\columnwidth]{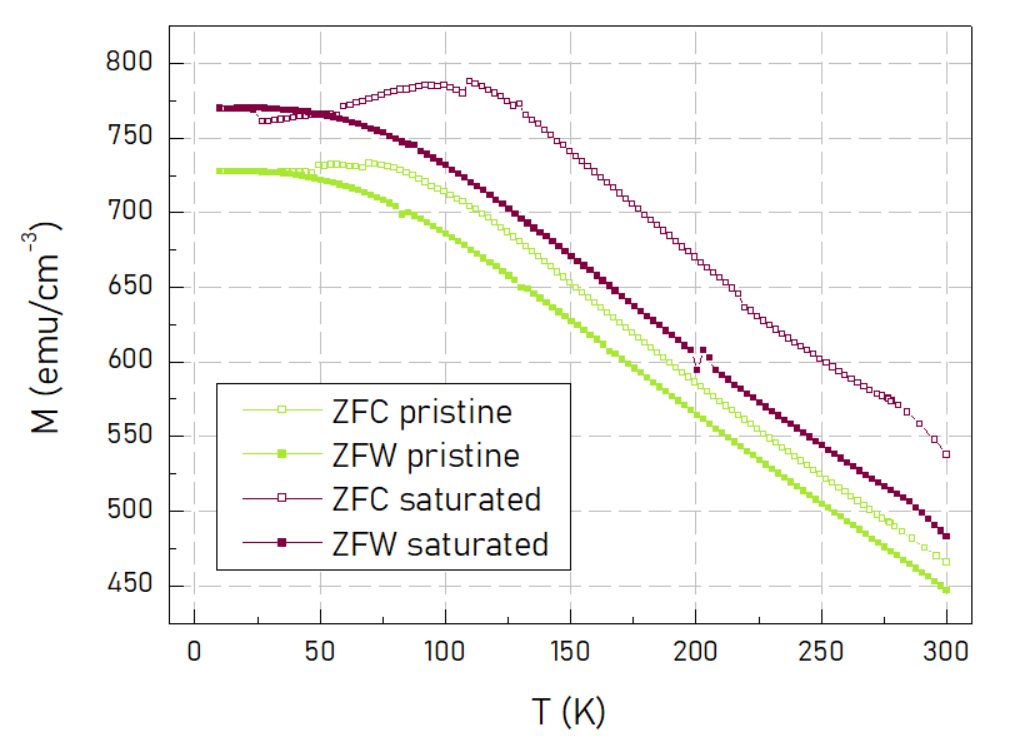}
\caption{\label{fig:MT_F3s}In-plane magnetization ($M$) as a function of temperature (T) recorded on sample \textbf{F45}. Two conditions are compared: without any specific treatment immediately before recording the loop (labeled as 'pristine') and in-plane saturated.} 
\end{figure}

Figure \ref{fig:MT_F3s} represents the remnant in-plane magnetization $M_{rem}$ measured on \textbf{F45} upon zero-field cooling down (ZFC) and warming up (ZFW). Two different initial conditions were set: pristine (i.e. without any special preparation beforehand) and saturated (i.e. having been subjected to 2 T applied in an in-plane configuration). On the one hand, $M^{saturated}_{rem}$ $>$ $M^{pristine}_{rem}$. This is reasonable considering that in the saturated case \textbf{M} is aligned with $H_{ext}$ meaning that all the in-plane components in the remnant state are contributing to each other. Alternatively, $M^{ZFC}_{rem}$ $>$ $M^{ZFW}_{rem}$ is understandable because the configuration upon cooling down is necessarily better ordered than warming up. This occurs because the system temperature gets reduced below the condition at which the stripes vanish and when they form again their relative orientation can be dissimilar. Upon lowering T, the increase in $M$ is related to the reduction of $K_{\rm PMA}$, which implies that a larger component of \textbf{M} lies in-plane.  

Consistent with the MFM image of \textbf{F60} at 150 K (see Fig. \ref{fig:MFM-images}), the characterization of \textbf{M} depends considerably on the thermal history and, specifically, on whether T is dropping or raising. This observation, which is easily understandable when considering that coming from a disordered less-anisotropic low-T configuration results in a poorly aligned striped pattern, highlights the need for including memory effects when modeling the magnetic pattern. This is discussed in the following section. 

\section*{\label{sec:Discussion}Discussion}

\noindent A closer inspection of the $M-H$ dependence (Fig. \ref{fig:MH_F1s_F3s}) offers a deeper insight into the phenomenology involved in the transition from the striped pattern to the fully-planar homogeneous magnetization. As previously mentioned, in the case of sample \textbf{F75}, an increase of $M_{rem}$ as well as of $H_C$ both agree with the same physical mechanism. Upon T reduction, the in-plane component of \textbf{M}  (\textbf{M}$_{IP}$ $\equiv M$) gets favored to the detriment of the out-of-plane one. 
%
%
This evolution can be expressed in terms of the so-called deflection angle ($\theta_0$), which is defined as the angle determined between the magnetization vector and the film plane. Although this picture assumes a very simple tridimensional structure of the domains, it comprises a useful phenomenological model to quantify the changes in an attempt for correlating different properties of the films \cite{saito_new_1964,martinez_modeling_2016}. Calculated as the inverse cosine of the ratio between $M_{rem}$ and $M_{sat}$, here we contrast it with the domain period, $\lambda$ (see Table \ref{tab2}). Extracted from the inverse of the 2D FFT periodicity (see Figs. S5 and S6 in the Supplementary Information), $\lambda$ can be roughly considered as two times the domain width. Since the two quantities, $\theta_0$ and $\lambda$ are, in turn, related to the competition between the exchange and dipolar interactions, a monotonous dependence between them is expected \cite{saito_new_1964}.

\begin{table}[htbp]
\caption{Maximum deflection angle, $\theta_0$, and domain period, $\lambda$, for two samples (\textbf{F75} and \textbf{F45}) at different T (300, 150, and 10 K).}
\begin{center}
\begin{tabular}{|c|c|c||c|c|c|}
\hline
\textbf{Sample} & T & \underline{$M_{rem}$} & $\theta_0$ & FFT maxima & $\lambda$ \\
 & [K] & $M_{sat}$ & [deg] & [$\mu m^{-1}$] & [nm] \\
\hline
\textbf{F75} & 300 & 0.4 & $\sim$66 & $\sim$4 & 250 \\
\hline
\textbf{F75} & 150 & 0.5 & $\sim$60 & $\sim$5 & 200 \\
\hline
\textbf{F75} & 10 & 0.6 & $\sim$53 & $\sim$6 & 167 \\
\hline
\textbf{F45} & 300 & 0.6 & $\sim$53 & $\sim$8 & 125 \\
\hline
\textbf{F45} & 150 & 0.7 & $\sim$46 & $\sim$3 & 333 \\
\hline
\textbf{F45} & 10 & 0.8 & $\sim$37 & - & \\
\hline
\end{tabular}
\label{tab2}
\end{center}
\end{table}

\noindent This is, in fact, the tendency observed for \textbf{F75}. On the contrary, in \textbf{F45} the domain width increases while the deflection angle gets reduced. 
%
%
Combined with the associated MFM images and $M-T$ loop, this observation accounts for \textbf{F45} being at the precise threshold at which, within the explored T-range, the thickness drops below the critical value. This explanation is supported by the maximum observed in the in-plane magnetization as a function of T (Fig. \ref{fig:MT_F3s}). Specifically, the T at which such a maximum is observed would indicate the condition at which the ordering, represented by the stripes, is macroscopically lost. The distortion does not occur immediately. The stripes progressively evolve into disconnected elongated domains which eventually lose the long-scale order. The observation of zig-zag patterns (see the condition F45 at 10 K in Fig. \ref{fig:MFM-images}) could be associated with the occurrence of smectic phases, as suggested in the literature \cite{seul_domain_1995}. Upon increasing T after having vanished the stripes, the maximum in \textit{M} is no longer reachable. Since it was the result of a highly ordered situation, it cannot be recovered unless an anisotropic stimulus breaks the symmetry to favor a preferential direction. The disconnected or zig-zag patterns observed for \textbf{F45} at 10 K recover by means of waiver stripes (illustrated by comparing Figs. S7 and S8, with S9 and S10) that overall give a smaller macroscopic in-plane magnetization due to internal mutual cancellations. This is true for both the pristine and saturated conditions, although the initial magnetization in the latter is obviously the highest. In this case, it is less likely that a simple model \cite{saito_new_1964,martinez_modeling_2016}, considering just two types of FM domains, still holds making it necessary to involve the tridimensional structure already recognized in other striped systems \cite{granada_magnetotransport_2016,camara_magnetization_2017,pianciola_magnetoresistance_2020}.          

It is also important to highlight that thermal cycling introduces spatial distortion in the striped pattern. The MFM images acquired in different zones of the same sample (see Supplementary Information) demonstrate that cycling T affects not only the domain width but also its lateral distribution. Even though mapping the stripes at RT before cycling T presents the same qualitative pattern, regardless of the specific point where the tip is landed (Figs. S7 and S8), repeating such a procedure after lowering T impacts the spatial dependence of the domain texture (see Figs. S9 and S10). %
%
Neighboring scanning areas display not only different orientations of the stripes but also changes in their periodicities. This, in turn, implies that the magnetic pattern obtained in a specific scanning window might not be representative of the whole sample (as it was before cycling T). Moreover, even when this phenomenology might be understandable for those samples such as \textbf{F45} that upon lowering T have crossed the critical thickness, this seems to apply to all the samples (as demonstrated by the fact that the MFM image of \textbf{F60} at 150 K seems to depend on whether T is lowering or raising). 

In addition, it is necessary to make a special mention of sample \textbf{F45}. Fig. \ref{fig:MH_F1s_F3s_b} shows the $M$ - $H$ loop shrinking from 300 K to 150 K even though Fig. \ref{fig:MT_F3s} seems to indicate a transition between 150 K and 10 K. The latter is in agreement with the MFM images. Nevertheless, the transition between stripes and in-plane magnetization is not sharp as already pointed out in the literature \cite{guzman_abnormal_2013} explaining the apparent contradiction. 

%

Finally, cycling T affects the balance among the magnetization free-energy terms in the same way as applying different magnetic fields. It is thus reasonable to expect the occurrence of metastable states, depending on the thermal history, as much as the sequence of the applied fields determines the magnetic state of the system and, in particular, its domain pattern. 

\section*{\label{sec:Conclusions}Conclusions}

\noindent The temperature evolution of the domain texture in A1 FePt thin films of various thicknesses has been studied. Low-temperature MFM images confirm that the critical thickness for the appearance of the striped pattern varies with temperature. The transition between ordered (striped) and disordered (in-plane) states is also revealed macroscopically in the behavior of the derivative of the magnetization in $M-H$ loops.

Additionally, by considering the relationship between the domain periodicity, $\lambda$, and T, we have validated at low T the dependence proposed for T $>>$ RT in the proximity of the Curie temperature. Moreover, we showed that varying T not only affects the periodicity of the stripes but also their mutual orientation. In particular, these changes appear to be local such that the formerly homogeneity in the in-plane magnetization is lost. Patches consisting of different magnetization orientations are observed. This is also reflected macroscopically in the difference in the temperature evolution of the magnetization upon cooling and warming the films. 

In summary, we have analyzed the distortion of the stripes in an archetypical FM system at low T, 
demonstrating that cycling T leads to metastable states that might not be properly described by a simplified modeled picture.

\begin{acknowledgments}
\noindent The authors would like to acknowledge 
the fruitful discussions held with Mara Granada, Alejandro Butera, and Beatriz Noheda.

\noindent C.P.Q. gratefully acknowledges financial support from EU-H2020-RISE project \textit{Memristive and multiferroic materials for logic units in nanoelectronics} 'MELON' (SEP-2106565560). D.G. and J.M. acknowledge ANPCyT funding via PICT 2018-01394.
\end{acknowledgments}




\bibliographystyle{apsrev4-1}
\bibliography{FePtStripes}

\begin{thebibliography}{30}%
\makeatletter
\providecommand \@ifxundefined [1]{%
 \@ifx{#1\undefined}
}%
\providecommand \@ifnum [1]{%
 \ifnum #1\expandafter \@firstoftwo
 \else \expandafter \@secondoftwo
 \fi
}%
\providecommand \@ifx [1]{%
 \ifx #1\expandafter \@firstoftwo
 \else \expandafter \@secondoftwo
 \fi
}%
\providecommand \natexlab [1]{#1}%
\providecommand \enquote  [1]{``#1''}%
\providecommand \bibnamefont  [1]{#1}%
\providecommand \bibfnamefont [1]{#1}%
\providecommand \citenamefont [1]{#1}%
\providecommand \href@noop [0]{\@secondoftwo}%
\providecommand \href [0]{\begingroup \@sanitize@url \@href}%
\providecommand \@href[1]{\@@startlink{#1}\@@href}%
\providecommand \@@href[1]{\endgroup#1\@@endlink}%
\providecommand \@sanitize@url [0]{\catcode `\\12\catcode `\$12\catcode
  `\&12\catcode `\#12\catcode `\^12\catcode `\_12\catcode `\%12\relax}%
\providecommand \@@startlink[1]{}%
\providecommand \@@endlink[0]{}%
\providecommand \url  [0]{\begingroup\@sanitize@url \@url }%
\providecommand \@url [1]{\endgroup\@href {#1}{\urlprefix }}%
\providecommand \urlprefix  [0]{URL }%
\providecommand \Eprint [0]{\href }%
\providecommand \doibase [0]{http://dx.doi.org/}%
\providecommand \selectlanguage [0]{\@gobble}%
\providecommand \bibinfo  [0]{\@secondoftwo}%
\providecommand \bibfield  [0]{\@secondoftwo}%
\providecommand \translation [1]{[#1]}%
\providecommand \BibitemOpen [0]{}%
\providecommand \bibitemStop [0]{}%
\providecommand \bibitemNoStop [0]{.\EOS\space}%
\providecommand \EOS [0]{\spacefactor3000\relax}%
\providecommand \BibitemShut  [1]{\csname bibitem#1\endcsname}%
\let\auto@bib@innerbib\@empty
\bibitem [{\citenamefont {Kittel}(1946)}]{kittel_theory_1946}%
  \BibitemOpen
  \bibfield  {author} {\bibinfo {author} {\bibfnamefont {C.}~\bibnamefont
  {Kittel}},\ }\href {\doibase 10.1103/PhysRev.70.965} {\bibfield  {journal}
  {\bibinfo  {journal} {Phys. Rev.}\ }\textbf {\bibinfo {volume} {70}},\
  \bibinfo {pages} {965} (\bibinfo {year} {1946})}\BibitemShut {NoStop}%
\bibitem [{\citenamefont {Williams}\ and\ \citenamefont
  {Sherwood}(1957)}]{williams_magnetic_1957}%
  \BibitemOpen
  \bibfield  {author} {\bibinfo {author} {\bibfnamefont {H.~J.}\ \bibnamefont
  {Williams}}\ and\ \bibinfo {author} {\bibfnamefont {R.~C.}\ \bibnamefont
  {Sherwood}},\ }\href {\doibase 10.1063/1.1722801} {\bibfield  {journal}
  {\bibinfo  {journal} {Journal of Applied Physics}\ }\textbf {\bibinfo
  {volume} {28}},\ \bibinfo {pages} {548} (\bibinfo {year} {1957})}\BibitemShut
  {NoStop}%
\bibitem [{\citenamefont {Saito}\ \emph {et~al.}(1964)\citenamefont {Saito},
  \citenamefont {Fujiwara},\ and\ \citenamefont {Sugita}}]{saito_new_1964}%
  \BibitemOpen
  \bibfield  {author} {\bibinfo {author} {\bibfnamefont {N.}~\bibnamefont
  {Saito}}, \bibinfo {author} {\bibfnamefont {H.}~\bibnamefont {Fujiwara}}, \
  and\ \bibinfo {author} {\bibfnamefont {Y.}~\bibnamefont {Sugita}},\ }\href
  {\doibase 10.1143/JPSJ.19.1116} {\bibfield  {journal} {\bibinfo  {journal}
  {J. Phys. Soc. Jpn.}\ }\textbf {\bibinfo {volume} {19}},\ \bibinfo {pages}
  {1116} (\bibinfo {year} {1964})}\BibitemShut {NoStop}%
\bibitem [{\citenamefont {Murayama}(1966)}]{murayama_micromagnetics_1966}%
  \BibitemOpen
  \bibfield  {author} {\bibinfo {author} {\bibfnamefont {Y.}~\bibnamefont
  {Murayama}},\ }\href {\doibase 10.1143/JPSJ.21.2253} {\bibfield  {journal}
  {\bibinfo  {journal} {J. Phys. Soc. Jpn.}\ }\textbf {\bibinfo {volume}
  {21}},\ \bibinfo {pages} {2253} (\bibinfo {year} {1966})}\BibitemShut
  {NoStop}%
\bibitem [{\citenamefont {Rieck}\ \emph {et~al.}(2023)\citenamefont {Rieck},
  \citenamefont {Cipollini}, \citenamefont {Salverda}, \citenamefont
  {Quinteros}, \citenamefont {Schomaker},\ and\ \citenamefont
  {Noheda}}]{rieck_ferroelastic_nodate}%
  \BibitemOpen
  \bibfield  {author} {\bibinfo {author} {\bibfnamefont {J.~L.}\ \bibnamefont
  {Rieck}}, \bibinfo {author} {\bibfnamefont {D.}~\bibnamefont {Cipollini}},
  \bibinfo {author} {\bibfnamefont {M.}~\bibnamefont {Salverda}}, \bibinfo
  {author} {\bibfnamefont {C.~P.}\ \bibnamefont {Quinteros}}, \bibinfo {author}
  {\bibfnamefont {L.~R.~B.}\ \bibnamefont {Schomaker}}, \ and\ \bibinfo
  {author} {\bibfnamefont {B.}~\bibnamefont {Noheda}},\ }\href {\doibase
  10.1002/aisy.202200292} {\bibfield  {journal} {\bibinfo  {journal} {Advanced
  Intelligent Systems}\ }\textbf {\bibinfo {volume} {5}},\ \bibinfo {pages}
  {2200292} (\bibinfo {year} {2023})}\BibitemShut {NoStop}%
\bibitem [{\citenamefont {Salje}(2010)}]{salje_multiferroic_2010}%
  \BibitemOpen
  \bibfield  {author} {\bibinfo {author} {\bibfnamefont {E.~K.~H.}\
  \bibnamefont {Salje}},\ }\href {\doibase 10.1002/cphc.200900943} {\bibfield
  {journal} {\bibinfo  {journal} {ChemPhysChem}\ }\textbf {\bibinfo {volume}
  {11}},\ \bibinfo {pages} {940} (\bibinfo {year} {2010})}\BibitemShut
  {NoStop}%
\bibitem [{\citenamefont {Catalan}\ \emph {et~al.}(2012)\citenamefont
  {Catalan}, \citenamefont {Seidel}, \citenamefont {Ramesh},\ and\
  \citenamefont {Scott}}]{catalan_domain_2012}%
  \BibitemOpen
  \bibfield  {author} {\bibinfo {author} {\bibfnamefont {G.}~\bibnamefont
  {Catalan}}, \bibinfo {author} {\bibfnamefont {J.}~\bibnamefont {Seidel}},
  \bibinfo {author} {\bibfnamefont {R.}~\bibnamefont {Ramesh}}, \ and\ \bibinfo
  {author} {\bibfnamefont {J.~F.}\ \bibnamefont {Scott}},\ }\href {\doibase
  10.1103/RevModPhys.84.119} {\bibfield  {journal} {\bibinfo  {journal} {Rev.
  Mod. Phys.}\ }\textbf {\bibinfo {volume} {84}},\ \bibinfo {pages} {119}
  (\bibinfo {year} {2012})}\BibitemShut {NoStop}%
\bibitem [{\citenamefont {Likharev}(2007)}]{likharev_hybrid_2007}%
  \BibitemOpen
  \bibfield  {author} {\bibinfo {author} {\bibfnamefont {K.~K.}\ \bibnamefont
  {Likharev}},\ }\href {\doibase 10.1116/1.2794060} {\bibfield  {journal}
  {\bibinfo  {journal} {Journal of Vacuum Science \& Technology B:
  Microelectronics and Nanometer Structures Processing, Measurement, and
  Phenomena}\ }\textbf {\bibinfo {volume} {25}},\ \bibinfo {pages} {2531}
  (\bibinfo {year} {2007})}\BibitemShut {NoStop}%
\bibitem [{\citenamefont {Feigl}\ \emph {et~al.}(2014)\citenamefont {Feigl},
  \citenamefont {Yudin}, \citenamefont {Stolichnov}, \citenamefont {Sluka},
  \citenamefont {Shapovalov}, \citenamefont {Mtebwa}, \citenamefont {Sandu},
  \citenamefont {Wei}, \citenamefont {Tagantsev},\ and\ \citenamefont
  {Setter}}]{feigl_controlled_2014}%
  \BibitemOpen
  \bibfield  {author} {\bibinfo {author} {\bibfnamefont {L.}~\bibnamefont
  {Feigl}}, \bibinfo {author} {\bibfnamefont {P.}~\bibnamefont {Yudin}},
  \bibinfo {author} {\bibfnamefont {I.}~\bibnamefont {Stolichnov}}, \bibinfo
  {author} {\bibfnamefont {T.}~\bibnamefont {Sluka}}, \bibinfo {author}
  {\bibfnamefont {K.}~\bibnamefont {Shapovalov}}, \bibinfo {author}
  {\bibfnamefont {M.}~\bibnamefont {Mtebwa}}, \bibinfo {author} {\bibfnamefont
  {C.~S.}\ \bibnamefont {Sandu}}, \bibinfo {author} {\bibfnamefont {X.-K.}\
  \bibnamefont {Wei}}, \bibinfo {author} {\bibfnamefont {A.~K.}\ \bibnamefont
  {Tagantsev}}, \ and\ \bibinfo {author} {\bibfnamefont {N.}~\bibnamefont
  {Setter}},\ }\href {\doibase 10.1038/ncomms5677} {\bibfield  {journal}
  {\bibinfo  {journal} {Nat Commun}\ }\textbf {\bibinfo {volume} {5}},\
  \bibinfo {pages} {4677} (\bibinfo {year} {2014})}\BibitemShut {NoStop}%
\bibitem [{\citenamefont {Nahas}\ \emph {et~al.}(2020)\citenamefont {Nahas},
  \citenamefont {Prokhorenko}, \citenamefont {Fischer}, \citenamefont {Xu},
  \citenamefont {Carr{\'e}t{\'e}ro}, \citenamefont {Prosandeev}, \citenamefont
  {Bibes}, \citenamefont {Fusil}, \citenamefont {Dkhil}, \citenamefont
  {Garcia},\ and\ \citenamefont {Bellaiche}}]{nahas_inverse_2020}%
  \BibitemOpen
  \bibfield  {author} {\bibinfo {author} {\bibfnamefont {Y.}~\bibnamefont
  {Nahas}}, \bibinfo {author} {\bibfnamefont {S.}~\bibnamefont {Prokhorenko}},
  \bibinfo {author} {\bibfnamefont {J.}~\bibnamefont {Fischer}}, \bibinfo
  {author} {\bibfnamefont {B.}~\bibnamefont {Xu}}, \bibinfo {author}
  {\bibfnamefont {C.}~\bibnamefont {Carr{\'e}t{\'e}ro}}, \bibinfo {author}
  {\bibfnamefont {S.}~\bibnamefont {Prosandeev}}, \bibinfo {author}
  {\bibfnamefont {M.}~\bibnamefont {Bibes}}, \bibinfo {author} {\bibfnamefont
  {S.}~\bibnamefont {Fusil}}, \bibinfo {author} {\bibfnamefont
  {B.}~\bibnamefont {Dkhil}}, \bibinfo {author} {\bibfnamefont
  {V.}~\bibnamefont {Garcia}}, \ and\ \bibinfo {author} {\bibfnamefont
  {L.}~\bibnamefont {Bellaiche}},\ }\href {\doibase 10.1038/s41586-019-1845-4}
  {\bibfield  {journal} {\bibinfo  {journal} {Nature}\ }\textbf {\bibinfo
  {volume} {577}},\ \bibinfo {pages} {47} (\bibinfo {year} {2020})}\BibitemShut
  {NoStop}%
\bibitem [{\citenamefont {Cahn}(1991)}]{cahn_binary_1991}%
  \BibitemOpen
  \bibfield  {author} {\bibinfo {author} {\bibfnamefont {R.~W.}\ \bibnamefont
  {Cahn}},\ }\href {\doibase 10.1002/adma.19910031215} {\bibfield  {journal}
  {\bibinfo  {journal} {Advanced Materials}\ }\textbf {\bibinfo {volume} {3}},\
  \bibinfo {pages} {628} (\bibinfo {year} {1991})}\BibitemShut {NoStop}%
\bibitem [{\citenamefont {Vasquez~Mansilla}\ \emph {et~al.}(2008)\citenamefont
  {Vasquez~Mansilla}, \citenamefont {Gomez},\ and\ \citenamefont
  {Butera}}]{Vasquez2008}%
  \BibitemOpen
  \bibfield  {author} {\bibinfo {author} {\bibfnamefont {M.}~\bibnamefont
  {Vasquez~Mansilla}}, \bibinfo {author} {\bibfnamefont {J.}~\bibnamefont
  {Gomez}}, \ and\ \bibinfo {author} {\bibfnamefont {A.}~\bibnamefont
  {Butera}},\ }\href {\doibase 10.1109/TMAG.2008.2001518} {\bibfield  {journal}
  {\bibinfo  {journal} {IEEE Transactions on Magnetics}\ }\textbf {\bibinfo
  {volume} {44}},\ \bibinfo {pages} {2883} (\bibinfo {year}
  {2008})}\BibitemShut {NoStop}%
\bibitem [{\citenamefont {Sallica~Leva}\ \emph {et~al.}(2010)\citenamefont
  {Sallica~Leva}, \citenamefont {Valente}, \citenamefont
  {Mart{\'i}nez~Tabares}, \citenamefont {V{\'a}squez~Mansilla}, \citenamefont
  {Roshdestwensky},\ and\ \citenamefont {Butera}}]{sallica_leva_magnetic_2010}%
  \BibitemOpen
  \bibfield  {author} {\bibinfo {author} {\bibfnamefont {E.}~\bibnamefont
  {Sallica~Leva}}, \bibinfo {author} {\bibfnamefont {R.~C.}\ \bibnamefont
  {Valente}}, \bibinfo {author} {\bibfnamefont {F.}~\bibnamefont
  {Mart{\'i}nez~Tabares}}, \bibinfo {author} {\bibfnamefont {M.}~\bibnamefont
  {V{\'a}squez~Mansilla}}, \bibinfo {author} {\bibfnamefont {S.}~\bibnamefont
  {Roshdestwensky}}, \ and\ \bibinfo {author} {\bibfnamefont {A.}~\bibnamefont
  {Butera}},\ }\href {\doibase 10.1103/PhysRevB.82.144410} {\bibfield
  {journal} {\bibinfo  {journal} {Phys. Rev. B}\ }\textbf {\bibinfo {volume}
  {82}},\ \bibinfo {pages} {144410} (\bibinfo {year} {2010})}\BibitemShut
  {NoStop}%
\bibitem [{\citenamefont {Mart{\'i}nez}\ \emph {et~al.}(2016)\citenamefont
  {Mart{\'i}nez}, \citenamefont {Milano}, \citenamefont {Eddrief},
  \citenamefont {Marangolo},\ and\ \citenamefont
  {Bustingorry}}]{martinez_modeling_2016}%
  \BibitemOpen
  \bibfield  {author} {\bibinfo {author} {\bibfnamefont {M.~D.~P.}\
  \bibnamefont {Mart{\'i}nez}}, \bibinfo {author} {\bibfnamefont
  {J.}~\bibnamefont {Milano}}, \bibinfo {author} {\bibfnamefont
  {M.}~\bibnamefont {Eddrief}}, \bibinfo {author} {\bibfnamefont
  {M.}~\bibnamefont {Marangolo}}, \ and\ \bibinfo {author} {\bibfnamefont
  {S.}~\bibnamefont {Bustingorry}},\ }\href {\doibase
  10.1088/0953-8984/28/13/136001} {\bibfield  {journal} {\bibinfo  {journal}
  {J. Phys.: Condens. Matter}\ }\textbf {\bibinfo {volume} {28}},\ \bibinfo
  {pages} {136001} (\bibinfo {year} {2016})}\BibitemShut {NoStop}%
\bibitem [{\citenamefont {Fin}\ \emph {et~al.}(2015)\citenamefont {Fin},
  \citenamefont {Tomasello}, \citenamefont {Bisero}, \citenamefont {Marangolo},
  \citenamefont {Sacchi}, \citenamefont {Popescu}, \citenamefont {Eddrief},
  \citenamefont {Hepburn}, \citenamefont {Finocchio}, \citenamefont
  {Carpentieri}, \citenamefont {Rettori}, \citenamefont {Pini},\ and\
  \citenamefont {Tacchi}}]{fin_-plane_2015}%
  \BibitemOpen
  \bibfield  {author} {\bibinfo {author} {\bibfnamefont {S.}~\bibnamefont
  {Fin}}, \bibinfo {author} {\bibfnamefont {R.}~\bibnamefont {Tomasello}},
  \bibinfo {author} {\bibfnamefont {D.}~\bibnamefont {Bisero}}, \bibinfo
  {author} {\bibfnamefont {M.}~\bibnamefont {Marangolo}}, \bibinfo {author}
  {\bibfnamefont {M.}~\bibnamefont {Sacchi}}, \bibinfo {author} {\bibfnamefont
  {H.}~\bibnamefont {Popescu}}, \bibinfo {author} {\bibfnamefont
  {M.}~\bibnamefont {Eddrief}}, \bibinfo {author} {\bibfnamefont
  {C.}~\bibnamefont {Hepburn}}, \bibinfo {author} {\bibfnamefont
  {G.}~\bibnamefont {Finocchio}}, \bibinfo {author} {\bibfnamefont
  {M.}~\bibnamefont {Carpentieri}}, \bibinfo {author} {\bibfnamefont
  {A.}~\bibnamefont {Rettori}}, \bibinfo {author} {\bibfnamefont {M.~G.}\
  \bibnamefont {Pini}}, \ and\ \bibinfo {author} {\bibfnamefont
  {S.}~\bibnamefont {Tacchi}},\ }\href {\doibase 10.1103/PhysRevB.92.224411}
  {\bibfield  {journal} {\bibinfo  {journal} {Physical Review B}\ }\textbf
  {\bibinfo {volume} {92}},\ \bibinfo {pages} {224411} (\bibinfo {year}
  {2015})}\BibitemShut {NoStop}%
\bibitem [{\citenamefont {Álvarez}\ \emph {et~al.}(2014)\citenamefont
  {Álvarez}, \citenamefont {Sallica~Leva}, \citenamefont {Valente},
  \citenamefont {Vásquez~Mansilla}, \citenamefont {Gómez}, \citenamefont
  {Milano},\ and\ \citenamefont {Butera}}]{alvarez_correlation_2014}%
  \BibitemOpen
  \bibfield  {author} {\bibinfo {author} {\bibfnamefont {N.}~\bibnamefont
  {Álvarez}}, \bibinfo {author} {\bibfnamefont {E.}~\bibnamefont
  {Sallica~Leva}}, \bibinfo {author} {\bibfnamefont {R.~C.}\ \bibnamefont
  {Valente}}, \bibinfo {author} {\bibfnamefont {M.}~\bibnamefont
  {Vásquez~Mansilla}}, \bibinfo {author} {\bibfnamefont {J.}~\bibnamefont
  {Gómez}}, \bibinfo {author} {\bibfnamefont {J.}~\bibnamefont {Milano}}, \
  and\ \bibinfo {author} {\bibfnamefont {A.}~\bibnamefont {Butera}},\ }\href
  {\doibase 10.1063/1.4866685} {\bibfield  {journal} {\bibinfo  {journal}
  {Journal of Applied Physics}\ }\textbf {\bibinfo {volume} {115}},\ \bibinfo
  {pages} {083907} (\bibinfo {year} {2014})}\BibitemShut {NoStop}%
\bibitem [{\citenamefont {Quinteros}\ \emph {et~al.}(2020)\citenamefont
  {Quinteros}, \citenamefont {Burgos}, \citenamefont {Albornoz}, \citenamefont
  {Gomez}, \citenamefont {Granell}, \citenamefont {Golmar}, \citenamefont
  {Ibarra}, \citenamefont {Bustingorry}, \citenamefont {Curiale},\ and\
  \citenamefont {Granada}}]{quinteros_impact_2020}%
  \BibitemOpen
  \bibfield  {author} {\bibinfo {author} {\bibfnamefont {C.~P.}\ \bibnamefont
  {Quinteros}}, \bibinfo {author} {\bibfnamefont {M.~J.~C.}\ \bibnamefont
  {Burgos}}, \bibinfo {author} {\bibfnamefont {L.~J.}\ \bibnamefont
  {Albornoz}}, \bibinfo {author} {\bibfnamefont {J.}~\bibnamefont {Gomez}},
  \bibinfo {author} {\bibfnamefont {P.}~\bibnamefont {Granell}}, \bibinfo
  {author} {\bibfnamefont {F.}~\bibnamefont {Golmar}}, \bibinfo {author}
  {\bibfnamefont {M.~L.}\ \bibnamefont {Ibarra}}, \bibinfo {author}
  {\bibfnamefont {S.}~\bibnamefont {Bustingorry}}, \bibinfo {author}
  {\bibfnamefont {J.}~\bibnamefont {Curiale}}, \ and\ \bibinfo {author}
  {\bibfnamefont {M.}~\bibnamefont {Granada}},\ }\href
  {http://iopscience.iop.org/10.1088/1361-6463/abb849} {\bibfield  {journal}
  {\bibinfo  {journal} {J. Phys. D: Appl. Phys.}\ } (\bibinfo {year}
  {2020})}\BibitemShut {NoStop}%
\bibitem [{\citenamefont {Nečas}\ and\ \citenamefont
  {Klapetek}(2012)}]{necas_gwyddion_2012}%
  \BibitemOpen
  \bibfield  {author} {\bibinfo {author} {\bibfnamefont {D.}~\bibnamefont
  {Nečas}}\ and\ \bibinfo {author} {\bibfnamefont {P.}~\bibnamefont
  {Klapetek}},\ }\href {\doibase 10.2478/s11534-011-0096-2} {\bibfield
  {journal} {\bibinfo  {journal} {Open Physics}\ }\textbf {\bibinfo {volume}
  {10}},\ \bibinfo {pages} {181} (\bibinfo {year} {2012})}\BibitemShut
  {NoStop}%
\bibitem [{\citenamefont {Guzmán}\ \emph {et~al.}(2013)\citenamefont
  {Guzmán}, \citenamefont {Álvarez}, \citenamefont {Salva}, \citenamefont
  {Vásquez~Mansilla}, \citenamefont {Gómez},\ and\ \citenamefont
  {Butera}}]{guzman_abnormal_2013}%
  \BibitemOpen
  \bibfield  {author} {\bibinfo {author} {\bibfnamefont {J.~M.}\ \bibnamefont
  {Guzmán}}, \bibinfo {author} {\bibfnamefont {N.}~\bibnamefont {Álvarez}},
  \bibinfo {author} {\bibfnamefont {H.~R.}\ \bibnamefont {Salva}}, \bibinfo
  {author} {\bibfnamefont {M.}~\bibnamefont {Vásquez~Mansilla}}, \bibinfo
  {author} {\bibfnamefont {J.}~\bibnamefont {Gómez}}, \ and\ \bibinfo {author}
  {\bibfnamefont {A.}~\bibnamefont {Butera}},\ }\href {\doibase
  10.1016/j.jmmm.2013.07.037} {\bibfield  {journal} {\bibinfo  {journal}
  {Journal of Magnetism and Magnetic Materials}\ }\textbf {\bibinfo {volume}
  {347}},\ \bibinfo {pages} {61} (\bibinfo {year} {2013})}\BibitemShut
  {NoStop}%
\bibitem [{\citenamefont {Álvarez}\ \emph {et~al.}(2015)\citenamefont
  {Álvarez}, \citenamefont {Montalbetti}, \citenamefont {Gómez},
  \citenamefont {Riffo}, \citenamefont {Álvarez}, \citenamefont {Goovaerts},\
  and\ \citenamefont {Butera}}]{alvarez_tunable_2015}%
  \BibitemOpen
  \bibfield  {author} {\bibinfo {author} {\bibfnamefont {N.~R.}\ \bibnamefont
  {Álvarez}}, \bibinfo {author} {\bibfnamefont {M.~E.~V.}\ \bibnamefont
  {Montalbetti}}, \bibinfo {author} {\bibfnamefont {J.~E.}\ \bibnamefont
  {Gómez}}, \bibinfo {author} {\bibfnamefont {A.~E.~M.}\ \bibnamefont
  {Riffo}}, \bibinfo {author} {\bibfnamefont {M.~A.~V.}\ \bibnamefont
  {Álvarez}}, \bibinfo {author} {\bibfnamefont {E.}~\bibnamefont {Goovaerts}},
  \ and\ \bibinfo {author} {\bibfnamefont {A.}~\bibnamefont {Butera}},\ }\href
  {\doibase 10.1088/0022-3727/48/40/405003} {\bibfield  {journal} {\bibinfo
  {journal} {Journal of Physics D: Applied Physics}\ }\textbf {\bibinfo
  {volume} {48}},\ \bibinfo {pages} {405003} (\bibinfo {year}
  {2015})}\BibitemShut {NoStop}%
\bibitem [{\citenamefont {Álvarez}\ \emph {et~al.}(2016)\citenamefont
  {Álvarez}, \citenamefont {Gómez}, \citenamefont {Moya~Riffo}, \citenamefont
  {Vicente~Álvarez},\ and\ \citenamefont {Butera}}]{alvarez_critical_2016}%
  \BibitemOpen
  \bibfield  {author} {\bibinfo {author} {\bibfnamefont {N.~R.}\ \bibnamefont
  {Álvarez}}, \bibinfo {author} {\bibfnamefont {J.~E.}\ \bibnamefont
  {Gómez}}, \bibinfo {author} {\bibfnamefont {A.~E.}\ \bibnamefont
  {Moya~Riffo}}, \bibinfo {author} {\bibfnamefont {M.~A.}\ \bibnamefont
  {Vicente~Álvarez}}, \ and\ \bibinfo {author} {\bibfnamefont
  {A.}~\bibnamefont {Butera}},\ }\href {\doibase 10.1063/1.4942652} {\bibfield
  {journal} {\bibinfo  {journal} {Journal of Applied Physics}\ }\textbf
  {\bibinfo {volume} {119}},\ \bibinfo {pages} {083906} (\bibinfo {year}
  {2016})}\BibitemShut {NoStop}%
\bibitem [{\citenamefont {Sharma}\ \emph {et~al.}(2011)\citenamefont {Sharma},
  \citenamefont {Kaushik}, \citenamefont {Makino},\ and\ \citenamefont
  {Inoue}}]{sharma_anomalous_2011}%
  \BibitemOpen
  \bibfield  {author} {\bibinfo {author} {\bibfnamefont {P.}~\bibnamefont
  {Sharma}}, \bibinfo {author} {\bibfnamefont {N.}~\bibnamefont {Kaushik}},
  \bibinfo {author} {\bibfnamefont {A.}~\bibnamefont {Makino}}, \ and\ \bibinfo
  {author} {\bibfnamefont {A.}~\bibnamefont {Inoue}},\ }\href {\doibase
  10.1109/TMAG.2011.2159366} {\bibfield  {journal} {\bibinfo  {journal} {IEEE
  Transactions on Magnetics}\ }\textbf {\bibinfo {volume} {47}},\ \bibinfo
  {pages} {4394} (\bibinfo {year} {2011})}\BibitemShut {NoStop}%
\bibitem [{\citenamefont {Berger}\ and\ \citenamefont
  {Hopster}(1996)}]{berger_magnetization_1996}%
  \BibitemOpen
  \bibfield  {author} {\bibinfo {author} {\bibfnamefont {A.}~\bibnamefont
  {Berger}}\ and\ \bibinfo {author} {\bibfnamefont {H.}~\bibnamefont
  {Hopster}},\ }\href {\doibase 10.1063/1.362261} {\bibfield  {journal}
  {\bibinfo  {journal} {Journal of Applied Physics}\ }\textbf {\bibinfo
  {volume} {79}},\ \bibinfo {pages} {5619} (\bibinfo {year}
  {1996})}\BibitemShut {NoStop}%
\bibitem [{\citenamefont {Barker}\ and\ \citenamefont
  {Gehring}(1983)}]{barker_dipolar_1983}%
  \BibitemOpen
  \bibfield  {author} {\bibinfo {author} {\bibfnamefont {W.~A.}\ \bibnamefont
  {Barker}}\ and\ \bibinfo {author} {\bibfnamefont {G.~A.}\ \bibnamefont
  {Gehring}},\ }\href {\doibase 10.1088/0022-3719/16/33/014} {\bibfield
  {journal} {\bibinfo  {journal} {J. Phys. C: Solid State Phys.}\ }\textbf
  {\bibinfo {volume} {16}},\ \bibinfo {pages} {6415} (\bibinfo {year}
  {1983})}\BibitemShut {NoStop}%
\bibitem [{\citenamefont {Seul}\ and\ \citenamefont
  {Wolfe}(1992)}]{seul_evolution_1992}%
  \BibitemOpen
  \bibfield  {author} {\bibinfo {author} {\bibfnamefont {M.}~\bibnamefont
  {Seul}}\ and\ \bibinfo {author} {\bibfnamefont {R.}~\bibnamefont {Wolfe}},\
  }\href {\doibase 10.1103/PhysRevA.46.7519} {\bibfield  {journal} {\bibinfo
  {journal} {Physical Review A}\ }\textbf {\bibinfo {volume} {46}},\ \bibinfo
  {pages} {7519} (\bibinfo {year} {1992})}\BibitemShut {NoStop}%
\bibitem [{\citenamefont {Seul}\ and\ \citenamefont
  {Andelman}(1995)}]{seul_domain_1995}%
  \BibitemOpen
  \bibfield  {author} {\bibinfo {author} {\bibfnamefont {M.}~\bibnamefont
  {Seul}}\ and\ \bibinfo {author} {\bibfnamefont {D.}~\bibnamefont
  {Andelman}},\ }\href {\doibase 10.1126/science.267.5197.476} {\bibfield
  {journal} {\bibinfo  {journal} {Science}\ }\textbf {\bibinfo {volume}
  {267}},\ \bibinfo {pages} {476} (\bibinfo {year} {1995})}\BibitemShut
  {NoStop}%
\bibitem [{\citenamefont {Granada}\ \emph {et~al.}(2016)\citenamefont
  {Granada}, \citenamefont {Bustingorry}, \citenamefont {Pontello},
  \citenamefont {Barturen}, \citenamefont {Eddrief}, \citenamefont
  {Marangolo},\ and\ \citenamefont {Milano}}]{granada_magnetotransport_2016}%
  \BibitemOpen
  \bibfield  {author} {\bibinfo {author} {\bibfnamefont {M.}~\bibnamefont
  {Granada}}, \bibinfo {author} {\bibfnamefont {S.}~\bibnamefont
  {Bustingorry}}, \bibinfo {author} {\bibfnamefont {D.~E.}\ \bibnamefont
  {Pontello}}, \bibinfo {author} {\bibfnamefont {M.}~\bibnamefont {Barturen}},
  \bibinfo {author} {\bibfnamefont {M.}~\bibnamefont {Eddrief}}, \bibinfo
  {author} {\bibfnamefont {M.}~\bibnamefont {Marangolo}}, \ and\ \bibinfo
  {author} {\bibfnamefont {J.}~\bibnamefont {Milano}},\ }\href {\doibase
  10.1103/PhysRevB.94.184435} {\bibfield  {journal} {\bibinfo  {journal} {Phys.
  Rev. B}\ }\textbf {\bibinfo {volume} {94}},\ \bibinfo {pages} {184435}
  (\bibinfo {year} {2016})}\BibitemShut {NoStop}%
\bibitem [{\citenamefont {Camara}\ \emph {et~al.}(2017)\citenamefont {Camara},
  \citenamefont {Tacchi}, \citenamefont {Garnier}, \citenamefont {Eddrief},
  \citenamefont {Fortuna}, \citenamefont {Carlotti},\ and\ \citenamefont
  {Marangolo}}]{camara_magnetization_2017}%
  \BibitemOpen
  \bibfield  {author} {\bibinfo {author} {\bibfnamefont {I.~S.}\ \bibnamefont
  {Camara}}, \bibinfo {author} {\bibfnamefont {S.}~\bibnamefont {Tacchi}},
  \bibinfo {author} {\bibfnamefont {L.-C.}\ \bibnamefont {Garnier}}, \bibinfo
  {author} {\bibfnamefont {M.}~\bibnamefont {Eddrief}}, \bibinfo {author}
  {\bibfnamefont {F.}~\bibnamefont {Fortuna}}, \bibinfo {author} {\bibfnamefont
  {G.}~\bibnamefont {Carlotti}}, \ and\ \bibinfo {author} {\bibfnamefont
  {M.}~\bibnamefont {Marangolo}},\ }\href {\doibase 10.1088/1361-648X/aa8f36}
  {\bibfield  {journal} {\bibinfo  {journal} {J. Phys.: Condens. Matter}\
  }\textbf {\bibinfo {volume} {29}},\ \bibinfo {pages} {465803} (\bibinfo
  {year} {2017})}\BibitemShut {NoStop}%
\bibitem [{\citenamefont {Pianciola}\ \emph {et~al.}(2020)\citenamefont
  {Pianciola}, \citenamefont {Flewett}, \citenamefont {De~Biasi}, \citenamefont
  {Hepburn}, \citenamefont {Lounis}, \citenamefont {V{\'a}squez-Mansilla},
  \citenamefont {Granada}, \citenamefont {Barturen}, \citenamefont {Eddrief},
  \citenamefont {Sacchi}, \citenamefont {Marangolo},\ and\ \citenamefont
  {Milano}}]{pianciola_magnetoresistance_2020}%
  \BibitemOpen
  \bibfield  {author} {\bibinfo {author} {\bibfnamefont {B.}~\bibnamefont
  {Pianciola}}, \bibinfo {author} {\bibfnamefont {S.}~\bibnamefont {Flewett}},
  \bibinfo {author} {\bibfnamefont {E.}~\bibnamefont {De~Biasi}}, \bibinfo
  {author} {\bibfnamefont {C.}~\bibnamefont {Hepburn}}, \bibinfo {author}
  {\bibfnamefont {L.}~\bibnamefont {Lounis}}, \bibinfo {author} {\bibfnamefont
  {M.}~\bibnamefont {V{\'a}squez-Mansilla}}, \bibinfo {author} {\bibfnamefont
  {M.}~\bibnamefont {Granada}}, \bibinfo {author} {\bibfnamefont
  {M.}~\bibnamefont {Barturen}}, \bibinfo {author} {\bibfnamefont
  {M.}~\bibnamefont {Eddrief}}, \bibinfo {author} {\bibfnamefont
  {M.}~\bibnamefont {Sacchi}}, \bibinfo {author} {\bibfnamefont
  {M.}~\bibnamefont {Marangolo}}, \ and\ \bibinfo {author} {\bibfnamefont
  {J.}~\bibnamefont {Milano}},\ }\href {\doibase 10.1103/PhysRevB.102.054438}
  {\bibfield  {journal} {\bibinfo  {journal} {Phys. Rev. B}\ }\textbf {\bibinfo
  {volume} {102}},\ \bibinfo {pages} {054438} (\bibinfo {year}
  {2020})}\BibitemShut {NoStop}%
\bibitem [{\citenamefont {Jain}(1988)}]{jain_fundamentals_1988}%
  \BibitemOpen
  \bibfield  {author} {\bibinfo {author} {\bibfnamefont {A.~K.}\ \bibnamefont
  {Jain}},\ }\href@noop {} {\emph {\bibinfo {title} {Fundamentals of {Digital}
  {Image} {Processing}}}},\ \bibinfo {edition} {1st}\ ed.\ (\bibinfo
  {publisher} {Pearson},\ \bibinfo {year} {1988})\BibitemShut {NoStop}%
\end{thebibliography}%

\pagebreak
\widetext
\begin{center}
\textbf{\large Evolution of ferromagnetic stripes in FePt thin films at low temperature \\ SUPPLEMENTARY INFORMATION}
\end{center}
\setcounter{equation}{0}
\setcounter{figure}{0}
\setcounter{table}{0}
\setcounter{page}{1}
\makeatletter
\renewcommand{\theequation}{S\arabic{equation}}
\renewcommand{\thefigure}{S\arabic{figure}}
\renewcommand{\bibnumfmt}[1]{[S#1]}

\section*{Macroscopic characterization}

\begin{figure*}[ht!]
    \centering
    \includegraphics[width=\textwidth]{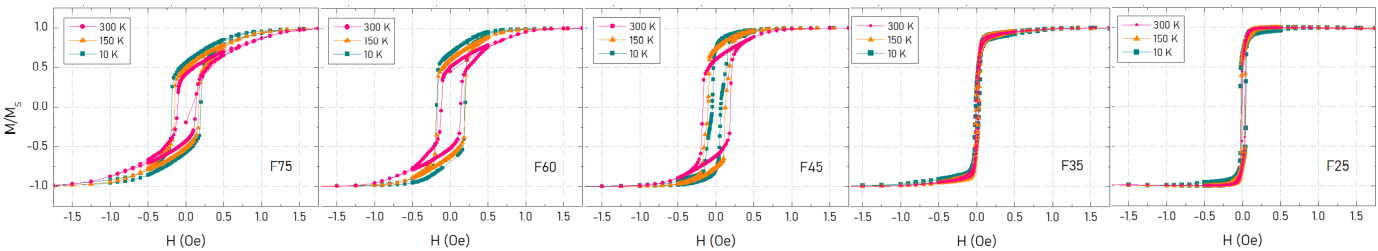}
    \caption{Normalized in-plane magnetization ($\frac{M}{M_{sat}}$) as a function of the magnetic field ($H$) acquired for each film at the three different T (300, 150, and 10 K).}
    \label{fig:MH_diffT}
\end{figure*}
\begin{figure*}[ht!]
    \centering
    \begin{subfigure}[b]{0.485\textwidth}
        \centering
        \includegraphics[width=\textwidth]{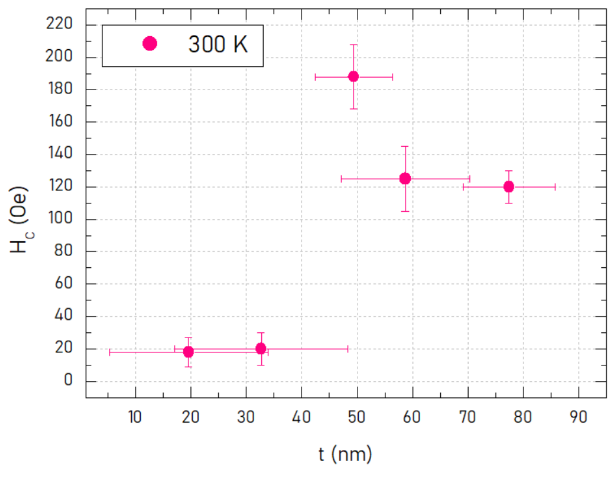}
        \caption{\label{fig:Hc-thick_RT}}
    \end{subfigure}%
    ~ 
    \begin{subfigure}[b]{0.5\textwidth}
        \centering
        \includegraphics[width=\textwidth]{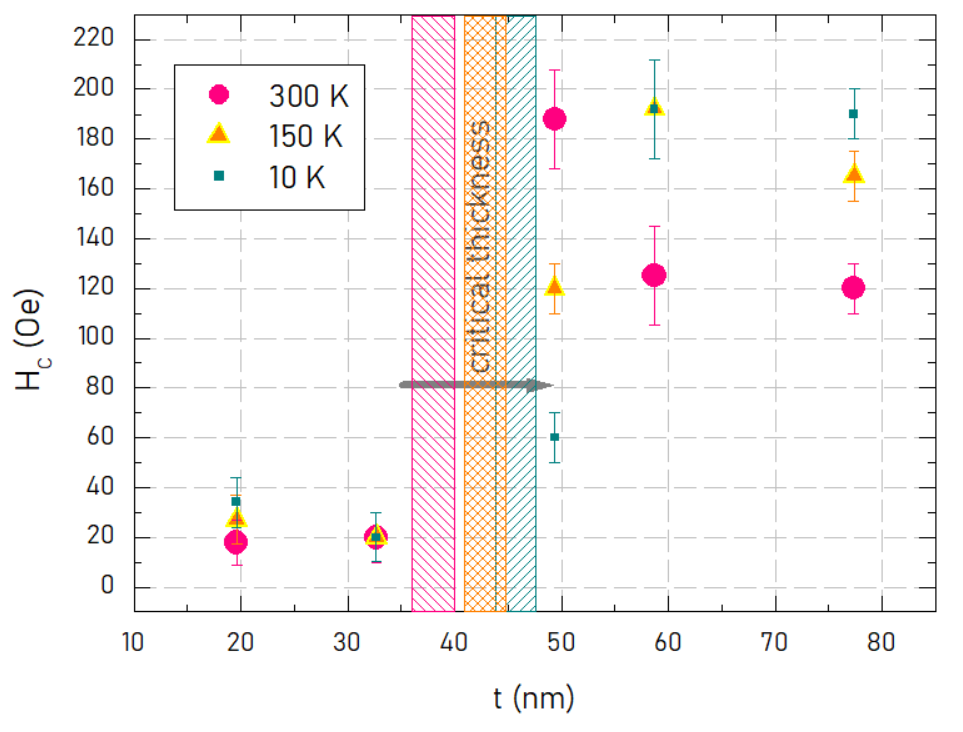}
        \caption{\label{fig:Hc-thick_diff-T}}
    \end{subfigure}
    \caption{\label{fig:Hc-tFePt_diff-T}Coercive field ($H_C$) as a function of film thickness ($t_{meas}$) (a) at RT and (b) at different T (300, 150, and 10 K). In the latter, each shadowed area represents an estimated range of the corresponding critical thickness ($t_{crit}$). The arrow indicates the increase of the critical thickness upon lowering T.} 
\end{figure*}
\begin{figure*}[ht!]
    \centering
    \begin{subfigure}[b]{0.49\textwidth}
        \centering
        \includegraphics[width=\textwidth]{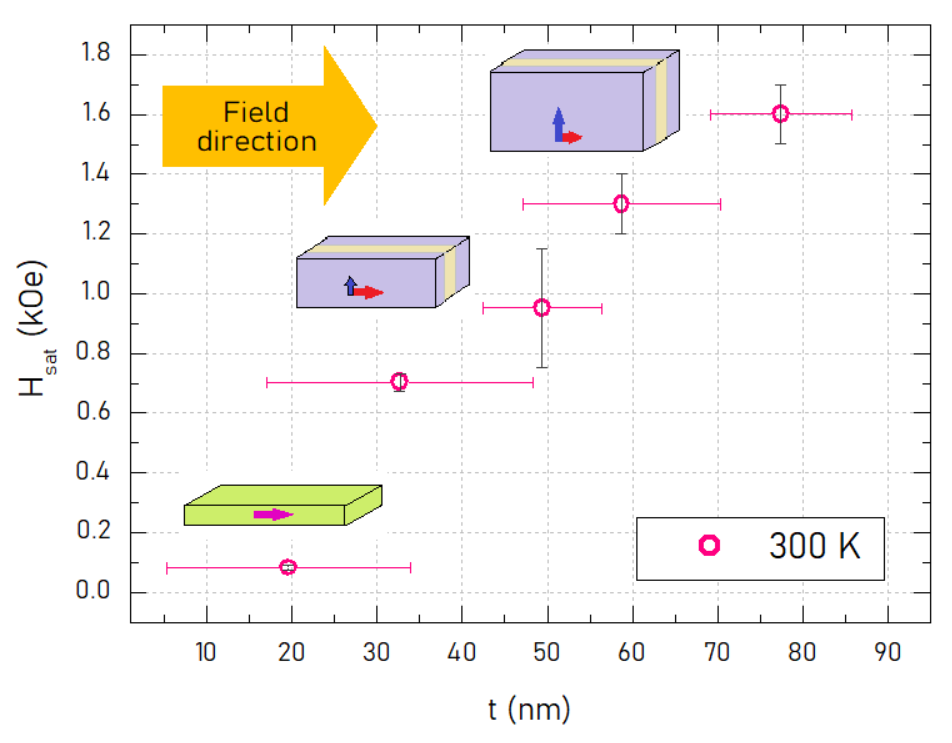}
        \caption{\label{fig:Hsat-thick_RT}}
    \end{subfigure}%
    ~ 
    \begin{subfigure}[b]{0.5\textwidth}
        \centering
        \includegraphics[width=\textwidth]{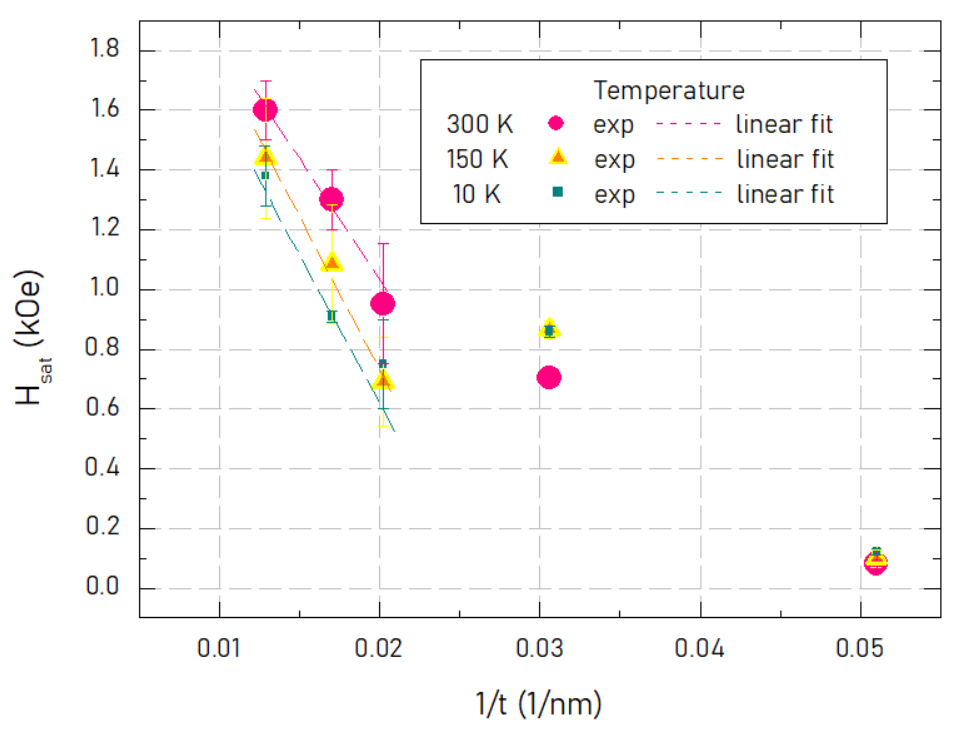}
        \caption{\label{fig:Hsat-invd_diff-T}}
    \end{subfigure}
    \caption{\label{fig:Hsat-tFePt_diff-T} (a) Saturation field ($H_{sat}$) as a function of film thickness ($t_{meas}$) at RT. The insets sketch the components of \textbf{M} depending on $t_{meas}$. (b) Saturation field ($H_{sat}$) as a function of the inverse of film thickness ($\frac{1}{t}$) at 300, 150, and 10 K. } 
\end{figure*}

\noindent In-plane magnetization (\textbf{M}$_{IP}$ $\equiv M$) was recorded as a function of externally applied in-plane magnetic field ($H$) for each FePt film at different T (see Fig. \ref{fig:MH_diffT}). From those $M-H$ loops, quantities such as the coercive field ($H_C$), the saturation field ($H_{sat}$), the saturation magnetization ($M_{sat}$), and the remnant magnetization ($M_{rem}$) were determined for the whole set of samples at different T. Figs. \ref{fig:Hc-tFePt_diff-T} and \ref{fig:Hsat-tFePt_diff-T} summarize the dependencies of $H_C$ and $H_{sat}$ as a function of the measured film thickness ($t$). 

\noindent Fig. \ref{fig:Hc-tFePt_diff-T}a depicts the evolution of $H_C$ with film thickness ($t_{meas}$) at RT. Fig. \ref{fig:Hc-tFePt_diff-T}b displays $H_C-t$ at different T. The coercive field ($H_C$) dramatically increases upon crossing $t_{crit}$ which, in turn, is a function of T, $t_{crit} \equiv f(\mathrm{T})$. For $t$ $>$ $t_{crit}$, a reduction in $H_C$ is associated with a further thickness increase. This can be understood considering $H_C$ as the condition for the already in-plane \textbf{M} component ($M$) to be reversed and it is consistent with previous results \cite{sallica_leva_magnetic_2010}. 


In turn, Fig. \ref{fig:Hsat-tFePt_diff-T}a depicts the evolution of $H_{sat}$ with $t_{meas}$ at RT. The insets sketch the magnetization components (\textbf{M}) for different films, including the reference to the direction of the external field. Fig. \ref{fig:Hsat-tFePt_diff-T}b represents $H_{sat}$ as a function of the inverse of $t_{meas}$ ($\frac{1}{t}$) at different T. 


The saturation field ($H_{sat}$) increases monotonously with the increase in thickness. For $t$ $<$ $t_{crit}$, $H_{sat}$ is relatively small as \textbf{M} lies almost entirely in-plane (although not completely, otherwise $H_C$ = $H_{sat}$ would be expected). For $t$ $>$ $t_{crit}$, $H_{sat}$ is a measure of the energy that it takes to rotate \textbf{M} until it becomes fully aligned with the applied external field ($H_{ext}$).

\break
\noindent Both, the $H_C$ as well as the $H_{sat}$ dependencies as a function of $t_{meas}$ are consistent with the picture of competition between in-plane and out-of-plane free-energy favorable terms. Below $t_{crit}$, which is, in turn, T-dependent, coercivity, and saturation fields are small due to the squareness of the associated $M-H$ loops (see Fig. \ref{fig:MH_diffT}). Beyond $t_{crit}$, stripes form, and an out-of-plane magnetization component is expected. 

\break
\section*{Stripes width}

\begin{figure}[ht!]
    \centering
    \includegraphics[width=0.5\textwidth]{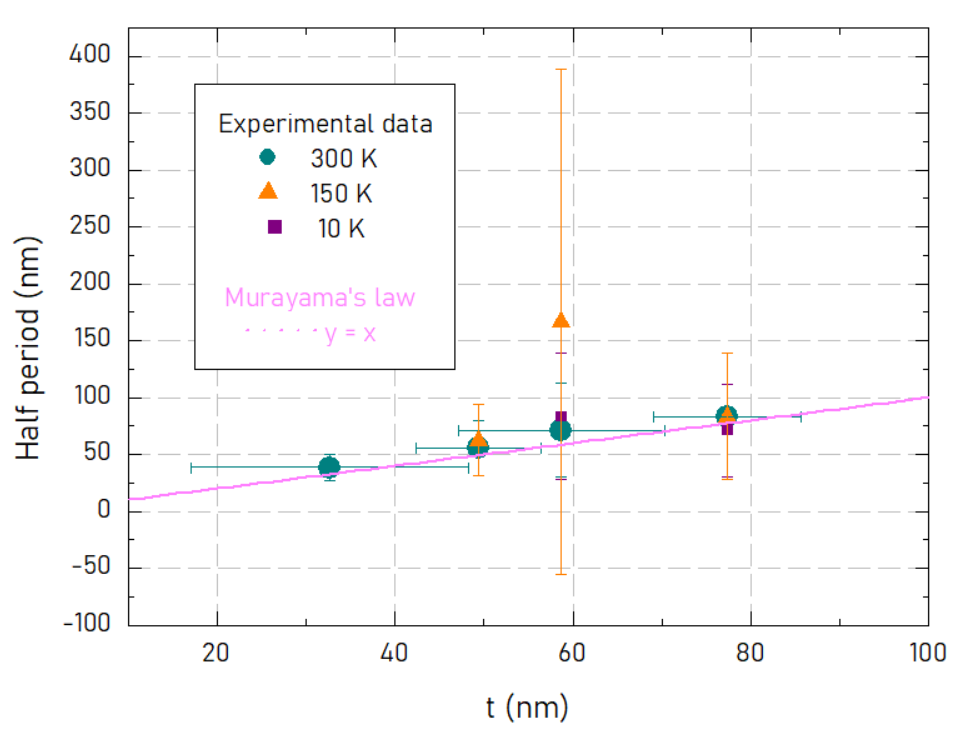}
    \caption{Stripes width as a function of $t_{meas}$ at three different T. Murayama's law is included as a reference. }
    \label{fig:Murayama}
\end{figure}

\noindent Considering the number of stripes contained in a certain unit length, it is possible to quantify the stripe width as half of their period. This strategy allows summarizing at a glance the information contained in Fig. 3 (of the main text). Fig. \ref{fig:Murayama} comprises the dependency of the stripes width as a function of $t_{meas}$ for different T. At RT (300 K), experimental data points agree with Murayama's law \cite{murayama_micromagnetics_1966}. For 150 K and 10 K, this is harder to say since the lower the T, the fewer films still display a striped pattern.   

\noindent Nevertheless, this plot does not distinguish among MFM images acquired while lowering or raising T and this has already been pointed out as a relevant aspect. The thermal history impacts the magnetic texture, as indicated in Fig. 3 and 4 of the main text. Not only does the periodicity get affected but even the possibility of assigning a representative value when a considerable dispersion is found both in the stripes period and orientation (including local misalignments). In the following, a transformation of the MFM image is used to consider not only the variation in the stripes period but also the widening of its associated uncertainty.    

\section*{2D FFT profiles}

\noindent Conducting the 2D Fast Fourier Transform (FFT) of the MFM images is a widely-used strategy to directly visualize the presence of periodicity. The observation of well-defined symmetric spots accounts for a marked periodicity and preferential orientation \cite{jain_fundamentals_1988}. Distortions from those highly-defined spots may arise either from the lack of preferential orientation or from the coexistence of more than one possible patch (zones with their own internal coherence). Moreover, the broadening and/or reduction of spots' intensity could indicate the appearance of multiple periodicities detected in the original image.  

\begin{figure}[ht!]
    \centering
    \includegraphics[width=0.5\textwidth]{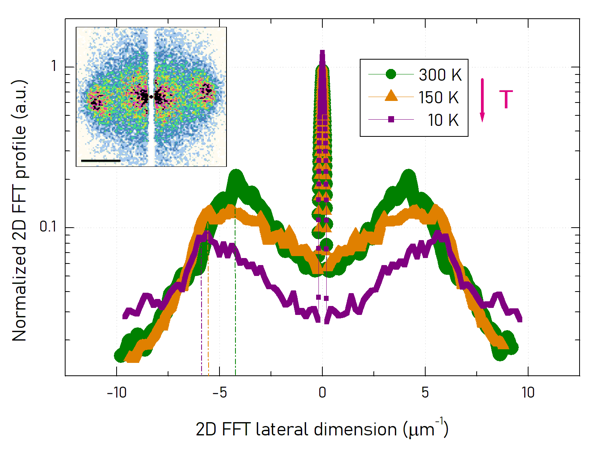}
    \caption{Profiles extracted from the 2D Fast Fourier Transformation (FFT) of the MFM images acquired on sample \textbf{F75} at three different T (300, 150, and 10 K). \underline{Inset}: 2D FFT of the RT (300 K) MFM measurement. The scale bar is 5 $\mu$m$^{-1}$-long.}
    \label{fig:2D-FFT_profile_F1_diff-T}
\end{figure}

\begin{figure}[ht!]
    \centering
    \includegraphics[width=0.5\textwidth]{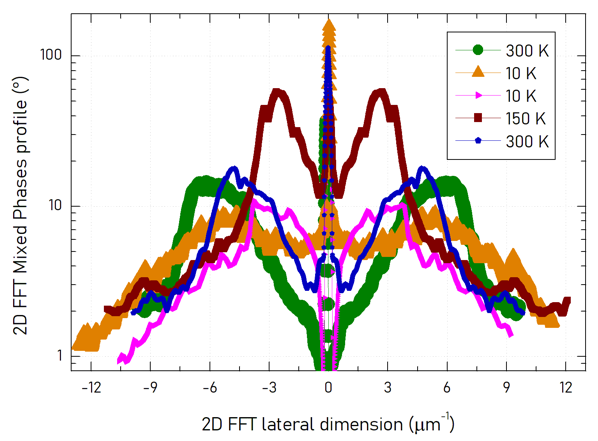}
    \caption{Profiles extracted from the 2D Fast Fourier Transformation (FFT) of the MFM images acquired on sample \textbf{F60} at three different T (300, 150, and 10 K).}
    \label{fig:2D-FFT_profile_F2_diff-T}
\end{figure}

The results of 2D FFT were used together with the MFM images as a visual tool to appreciate similarities and differences (see Fig. 3 of the main text). On the one hand, it is worth saying that each MFM has been rotated such that the stripes lie mainly in the vertical orientation. In a perfectly periodic, highly anisotropic situation, two intense spots are expected. On the contrary, the distortion from such a situation is observed in many of the experiments reported here. Nevertheless, the 2D FFT persists as the best tool for assigning a value to the stripes period. To perform such a task, we have extracted an average profile from them. Such a profile displays two relative maxima in the Fourier-transformed space from which the domain period can be extracted. Figs. \ref{fig:2D-FFT_profile_F1_diff-T} and \ref{fig:2D-FFT_profile_F2_diff-T} comprise the profiles extracted from the 2D FFTs of different films subjected to multiple T. 

Fig. \ref{fig:2D-FFT_profile_F1_diff-T} includes the profile of the 2D FFT of the MFM measured in \textbf{F75} while reducing T at three different conditions. Besides the maximum centered at 0 $\mu m^{-1}$, here we focus on the two relative maxima. 
The position of this reciprocal space, at which the maxima occur, is associated with the underlying periodicity, $\lambda$. Domain width would then be approximately half of this quantity. 

Fig. \ref{fig:2D-FFT_profile_F2_diff-T} depicts the profiles of the 2D FFT corresponding to the MFM images acquired on \textbf{F60} while varying T. By comparing the initial profile (at RT) and the one at 10 K, the spacing determined by the maxima gets reduced. This is the opposite trend compared to sample \textbf{F75}, see Fig. \ref{fig:2D-FFT_profile_F1_diff-T}. Additionally, the two profiles recorded at 300 K differ from each other. This could be either because lowering or increasing T determines different periodicities, due to the spatial dispersion observed after reducing T (which, in turn, implies that a certain scanning window might not be representative of the whole film), or both. This aspect is further discussed in the next section. 

\break
\section*{Lateral uniformity}

\subsection*{Before T-cycling}

\begin{figure*}[ht!]
    \centering
    \includegraphics[width=\textwidth]{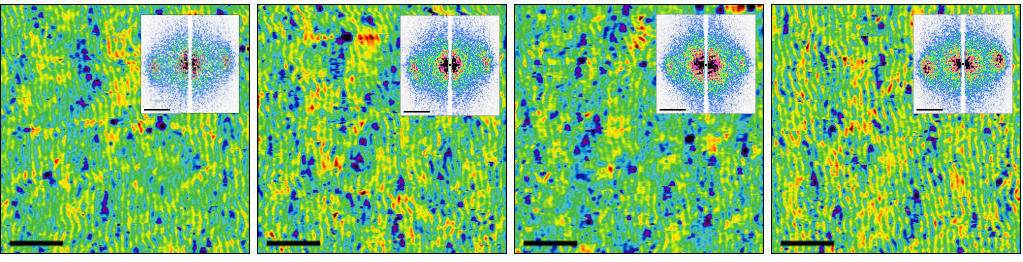}
    \caption{MFM images taken on the four corners of the sample \textbf{F75} (75 nm-thick FePt) to test the lateral uniformity. During acquisition, all the parameters (including the scan angle) were held constant. The scale bar is 1 $\mu$m-long. \underline{Insets}: 2D FFT of each associated MFM image. The scale bar is 5 $\mu$m$^{-1}$-long.}
    \label{fig:MFM_RT_F1}
\end{figure*}

\begin{figure*}[ht!]
    \centering
    \includegraphics[width=\textwidth]{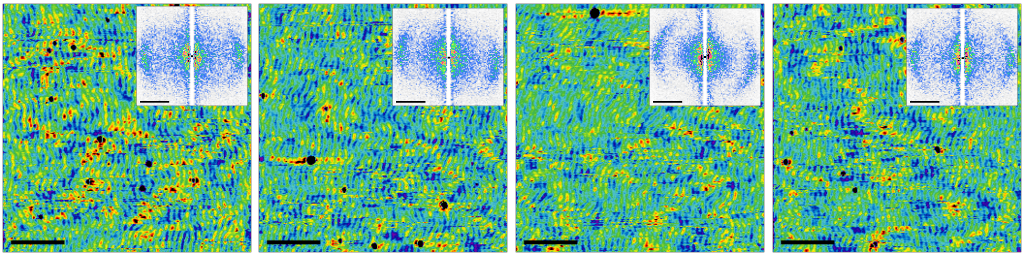}
    \caption{MFM images taken on the four corners of the sample \textbf{F45} to test the lateral uniformity. During acquisition, all the parameters (including the scan angle) were held constant. The scale bar is 1 $\mu$m-long. Although compared to the thicker film (\textbf{F75}, Fig. \ref{fig:2D-FFT_profile_F1_diff-T}), the stripes are wavier, still, the patterns of the four considered spots can be considered as an indication of lateral uniformity. \underline{Insets}: 2D FFT of each associated MFM image. The scale bar is 5 $\mu$m$^{-1}$-long.}
    \label{fig:MFM_RT_F3}
\end{figure*}

\noindent Before cycling the samples at low T, multiple zones of the samples were recorded, see Figs. \ref{fig:MFM_RT_F1} and \ref{fig:MFM_RT_F3}. Overall, the MFM images and the 2D FFT can be considered equivalent even though the scans were recorded in the four furthest corners of each sample (\textbf{F75} and \textbf{F45}, respectively). This is particularly relevant since in the measurements performed while varying T, the tip has to be disengaged before setting a different T value. For this reason, and due to the inherent limitations of the experimental setup, the landing point would hardly be exactly the same. Obtaining similar magnetic textures in different areas of each sample seems to indicate that any of them would be representative of the whole film. 

\break
\subsection*{During T-cycling}

\begin{figure*}[ht!]
    \centering
    \includegraphics[width=\textwidth]{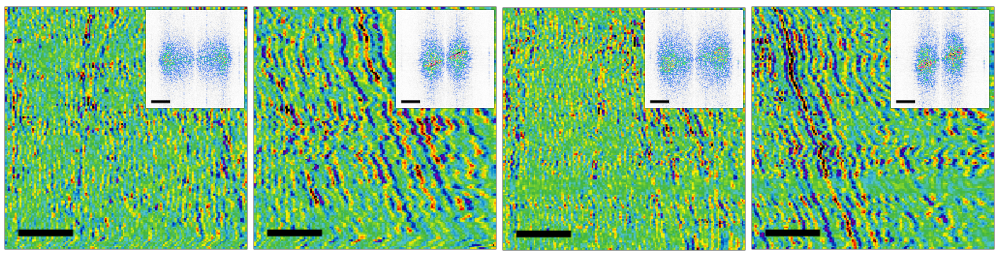}
    \caption{Spatial distribution of the stripes measured in the \textbf{F45} sample saturated at 300 K before cycling at low T. The four images represent MFM measurements acquired in four neighboring windows of (5 $\mu$m)$^2$-size. The insets display the 2D FFT of each window. All the parameters have been held constant to acquire them demonstrating that not only the periodicity but the orientation of the stripes is not uniform alongside the sample. The scale bar is 1 $\mu$m-long. \underline{Insets}: 2D FFT of each associated MFM image. The scale bar is 5 $\mu$m$^{-1}$-long.}
    \label{fig:MFM_RT_F3_decreasing-T_cycling}
\end{figure*}

\begin{figure*}[ht!]
    \centering
    \includegraphics[width=\textwidth]{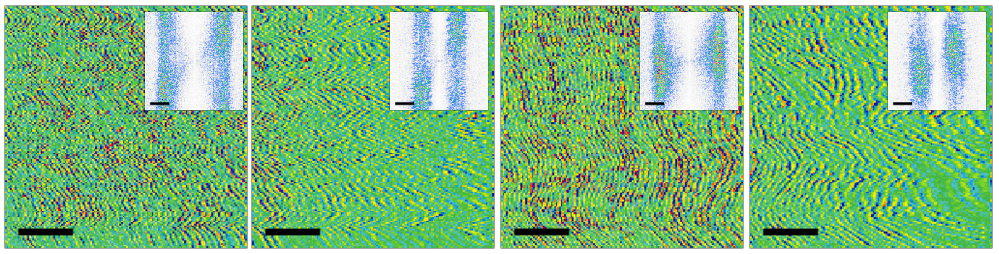}
    \caption{Spatial distribution of the stripes measured in the \textbf{F45} sample at 300 K after ZFW (see Fig. 4 of the main text). The four images represent MFM measurements acquired in four neighboring windows of (5 $\mu$m)$^2$-size. The insets display the 2D FFT of each window. While recording them, all the parameters have been held constant to demonstrate that not only the periodicity but the orientation of the stripes is dissimilar alongside the sample. Compared to Fig. \ref{fig:MFM_RT_F3_decreasing-T_cycling}, the stripes are even wavier than before indicating that a smaller net magnetization could be expected when macroscopically quantifying it. This is consistent with the difference in \textit{M} observed in Fig. 4 of the manuscript. The scale bar is 1 $\mu$m-long. \underline{Insets}: 2D FFT of each associated MFM image. The scale bar is 5 $\mu$m$^{-1}$-long.} %
    \label{fig:MFM_RT_F3_increasing-T_cycling}
\end{figure*}

\noindent During cycling at low T, we notice that the observation regarding the spatial homogeneity that we seemed to have found before, is not applicable. Scanning different areas of the same sample may produce completely different MFM images, as indicated by Figs. \ref{fig:MFM_RT_F3_decreasing-T_cycling} and \ref{fig:MFM_RT_F3_increasing-T_cycling}. 

Fig. \ref{fig:MFM_RT_F3_decreasing-T_cycling} consists of four MFM images of sample \textbf{F45} measured at RT immediately after saturating the sample. Fig. \ref{fig:MFM_RT_F3_increasing-T_cycling} also depicts four MFM images acquired at RT but after the sample has been cooled down to 10 K and then warmed up again. The two figures indicate that reiterative T cycling affects the stripes' periodicity and mutual orientation.  

\end{document}